\documentclass[12pt]{article}
\usepackage{latexsym}
\usepackage[DIV13]{typearea}
\usepackage{amsmath}
\usepackage{amssymb}
\usepackage{amsfonts}
\usepackage{cite}
\usepackage{bbold}
\usepackage[footnotesize]{caption2}
\usepackage{graphicx}
\usepackage[center,footnotesize,hang]{subfigure}
\usepackage{url}
\usepackage{lscape}
\usepackage{longtable}

\allowdisplaybreaks[1]

\newcommand{\eV}{\ensuremath{\,\mathrm{eV}}}

\newcommand{\GeV}{\ensuremath{\,\mathrm{GeV}}}
\newcommand{\TeV}{\ensuremath{\,\mathrm{TeV}}}
\newcommand{\be}{\begin{equation}}
\newcommand{\ee}{\end{equation}}
\newcommand{\order}[1]{\mathcal{O}\left(#1\right)}
\newcommand{\hc}{\ensuremath{\mathrm{h.c.}}}

\def\SU{\mathrm{SU}}
\def\U{\mathrm{U}}
\def\SO{\mathrm{SO}}

\DeclareMathOperator{\diag}{diag}

\newcommand{\Cl}[1]{\mathcal{C} _{#1}}

\newcommand{\Ord}[2]{\; ^{\circ} \mathrm{#1}_{#2}  \;}
\newcommand{\OrdCl}[1]{\; ^{\circ} \mathcal{C} _{#1} \;}
\newcommand{\Rep}[1]{\ensuremath{\underline{\mathbf{#1}}}}
\newcommand{\MoreRep}[2]{\ensuremath{\underline{\mathbf{#1}}_{\mathbf{#2}}}}
\newcommand{\Groupname}[2]{$ {#1} _{#2} $}

\newcommand{\Order}[1]{\ensuremath{\mathcal{O}\left(#1\right)}}

\newcommand{\Eqref}[1]{Eq.(\ref{#1})}

\newcommand{\Tabref}[1]{the Table \ref{#1}}

\newcommand{\Secref}[1]{Section \ref{#1}}
\newcommand{\Appref}[1]{the Appendix \ref{#1}}

\newcommand{\braket}[1]{\ensuremath{\left<#1\right>}}

\newcommand{\XX}{\ensuremath{\tilde A +2 \tilde B}}
\newcommand{\YY}{\ensuremath{\tilde A - \tilde B}}

\begin{document}
\begin{titlepage}
\ \vspace*{-15mm}
\begin{flushright}
IPPP/08/81\\
DCPT/08/162
\end{flushright}
\vspace*{5mm}

\begin{center}
\mathversion{bold}
{\LARGE \bf Lepton Mixing and Cancellation of the Dirac\\
Mass Hierarchy in SO(10) GUTs with\\ Flavor Symmetries $T_7$ and $\Sigma(81)$}\\
\mathversion{normal}

\vskip 0.5cm

{Claudia Hagedorn$^{a}$~\footnote{hagedorn@mpi-hd.mpg.de},
Michael A.~Schmidt$^{a,b}$~\footnote{m.a.schmidt@durham.ac.uk},
Alexei Yu.~Smirnov$^{a,c,d}$~\footnote{smirnov@ictp.it}
}
\\

\vskip 0.2cm
\setcounter{footnote}{0}

{\it $^a$ Max-Planck-Institut f\"{u}r Kernphysik, Postfach 10 39 80,\\ D-69029 Heidelberg
Germany\\
$^b$ Institute for Particle Physics Phenomenology, University of Durham,\\ Durham, DH1 3LE, UK\\ 
$^c$ International Centre for Theoretical Physics,
Strada Costiera 11,\\ 34014 Trieste, Italy\\
$^d$ Institute for Nuclear Research, RAS, Moscow, Russia
}
\end{center}

\abstract{
In $\SO(10)$ grand unified theories (GUTs)
the hierarchy which is present in the Dirac mass term of the neutrinos 
is  generically as strong as the one 
in the up-type quark mass term.
We propose a mechanism to partially or completely cancel 
this hierarchy in the light neutrino mass matrix  
in the seesaw context. 
The two main ingredients of the cancellation mechanism are
the existence of  three fermionic gauge singlets and of a discrete flavor symmetry $G_f$
which is broken at a higher scale than $\SO(10)$.
Two realizations of the cancellation mechanism are presented.
The realization 
based on the Frobenius group $T_7 \simeq Z_7\rtimes Z_3$ 
leads to a partial cancellation of the hierarchy 
and  relates maximal $2-3$ lepton mixing 
with the geometric hierarchy of the up-quark masses.
In the realization with the group $\Sigma(81)$ the cancellation is complete and tri-bimaximal lepton mixing is reproduced at the lowest order.
In both cases, to fully accommodate the leptonic data we take into account additional effects such as effects of
higher-dimensional operators involving more than one flavon.
The heavy neutral fermion mass spectra are considered.
For both realizations 
we analyze the flavon potential at the renormalizable level as well as ways to generate
the Cabibbo angle.
}
\end{titlepage}

\section{Introduction}
\label{sec:introduction}

The electric charge quantization as well as the possible gauge coupling unification
 at high energies are strong hints for a GUT~\cite{GUT}.
Especially an $\SO(10)$ GUT~\cite{SO10} looks very appealing, since
it allows one to unify all fermions of one generation including
the right-handed neutrino, $N \equiv  (\nu_R)^c$, into a single representation  $\Rep{16}$.
However, it is difficult to reconcile
this unification with the observation of a strong
hierarchy among the charged fermion masses, but only a mild one among
the neutrino masses. Indeed, a salient feature of the simplest versions 
of an $\SO(10)$ GUT is that the Dirac mass matrix $m_D$ of the neutrinos
has the same structure as the mass matrix of the up-quarks, 
{\it i.e.} it is strongly hierarchical.
In the type-I seesaw mechanism~\cite{seesaw} this matrix
appears twice and thereby, in general, leads to an even stronger  hierarchy among the light neutrino
masses contradicting observations. 
Furthermore, the diverse mixing patterns of quarks and leptons have to be explained.
It seems that the lepton sector reveals special features in
the mixings, such as $\mu-\tau$ symmetry~\cite{mutau} or tri-bimaximal mixing (TBM)~\cite{HPS}.
TBM, for example, can be understood in non-unified models
with the help of discrete or continuous flavor symmetries,  
such as $A_4$~\cite{A4tbm} and $\Delta(27)$~\cite{Delta27tbm} or $\SO(3)$~\cite{SO3tbm} and  $\SU(3)$~\cite{SU3tbm}.
In general this requires, however, that the fermions residing in different representations of 
the Standard Model (SM) gauge group, such as left- and right-handed components,
transform differently under $G_f$,  so that an extension of these models to
a GUT is not straightforward. There are several recent attempts to 
resolve this problem~\cite{Morisi,Stech,AFHA4,MalinskyKing}.

In this paper we propose a mechanism  to break the strong
correlation of the up-quark and the neutrino mass matrix in $\SO(10)$  
\footnote{Other approaches to this problem
can be found in~\cite{democraticapproach,Dermisek:2004tx,Sayre:2007ps}.}.
For this purpose, we assume  the existence of additional
fermionic GUT singlets $S_{i}$ and a discrete group $G_f$ to
constrain and correlate different couplings. 
The fields $S_{i}$ can mix with neutrinos only 
and thereby lead to  different properties of quark and
lepton mixings as well as to the smallness of neutrino masses 
~\cite{Smirnov:1993af,Smirnov:2004hs,Barr:2003nn,Lindner:2005pk,DiracScreening}. 
Each of the fermion generations is
accompanied by one singlet $S_{i}$. 
In our context, the mass matrix of the neutral fermions in the basis
$(\nu_L, N, S)$, is of the following form
\begin{equation}
\left( \nu_{L}, N, S \right) \, \left(
  \begin{array}{ccc}
    0 & m_{D}  & m_{\nu S}\\
    m_{D} ^{T} & 0 & M_{NS}\\
    m_{\nu S} ^{T} & M_{NS} ^{T} & M_{SS}
    \end{array}
\right) \, \left( \begin{array}{c} \nu_{L} \\ N \\ S \end{array} \right).
\label{neutrals}
\end{equation}
Block diagonalization of the matrix in \Eqref{neutrals}
yields the effective light neutrino mass matrix
\begin{equation}
m_{\nu}  \approx  m_{\nu}^{DS} + m_{\nu}^{LS},
\end{equation}
where
\begin{equation}
\label{eq:mnueffform}
m_\nu^{DS} =  m_D^{} \, (M_{NS}^{-1\, T} \, M_{SS}^{} \, M_{NS}^{-1}) \, m_D^T
\end{equation}
is the double seesaw (DS) contribution~\cite{DS}, and
\begin{equation}
m_{\nu}^{LS} = - \left[  m_{D}  \, \left( m_{\nu \, S} \, M_{N \, S} ^{-1}
\right) ^{T} + \left( m_{\nu \, S} \, M_{N \, S} ^{-1}
\right) \, m_{D} ^{T} \right]
\label{effmat}
\end{equation}
is the so-called linear seesaw (LS) contribution to the mass matrix~\cite{Barr:2003nn}.
If $m_{\nu}^{LS} \ll m_{\nu}^{DS}$ 
\footnote{This hierarchy arises, for instance, if $m_{D}$, $m_{\nu S}$ are of the order of 
the weak scale, $M_{NS}$ is of the order of the GUT scale  
and the masses of the SO(10) singlets are around the Planck scale.},   
the main contribution can be written as
\begin{equation}
\label{eq:mnueffform2}
m_\nu  \approx  m_\nu^{DS} =  F M_{SS}^{} F^T \; ,
\end{equation}
where
\begin{equation}
F \equiv m_D^{} M_{NS}^{-1\, T}.
\end{equation}

If the hierarchy present in the Dirac mass matrix $m_{D}$
is cancelled by the same or a similar hierarchy in $M_{NS}$, 
$F$ may turn out to be proportional to a matrix with $\mathcal{O}(1)$ entries.
Then the structure of the light neutrino mass matrix $m_{\nu}$ is non-hierarchical, 
provided that there is no hierarchy in the Majorana mass matrix $M_{SS}$ of the singlets $S_{i}$. 
We call this {\it complete cancellation} of the (Dirac mass) hierarchy. 
We refer to {\it partial cancellation}, if $F$ still contains some hierarchy.  
These possibilities are actually a generalization of the requirement $F \propto \mathbb{1}$
which arises if $M_{NS}^{T}$ is proportional to 
$m_{D}$~\cite{Smirnov:1993af,Smirnov:2004hs,Lindner:2005pk}.  
In this case, called the Dirac screening mechanism~\cite{Lindner:2005pk}, 
the light neutrino mass matrix is proportional to $M_{SS}$ and therefore
neutrino masses and lepton mixings are completely decoupled from charged fermion masses and quark mixings.
In the context of the type-I seesaw mechanism a similar cancellation has been
presented in~\cite{King}.

In this paper we show how the cancellation mechanism
can be realized in models with the discrete flavor symmetries $T_{7}$~\cite{Luhn:2007sy}
and $\Sigma(81)$~\cite{Ma:2006ht,Ma:2007ku}. 
The $T_{7}$ realization
leads to maximal atmospheric mixing and very small $\theta_{13}$
through a partial cancellation of the up-quark mass hierarchy in the neutrino sector.
The large value of the solar mixing angle $\theta_{12}$ 
however cannot be explained in this way and also not
through the effects of higher-dimensional operators involving more than one flavon. To generate large 
$\theta_{12}$ we have to
introduce an additional Higgs field. 
It only contributes to the LS term,
while not disturbing the partial cancellation arising in the DS contribution.
In the $\Sigma(81)$ setup the up-quark mass hierarchy is completely cancelled and
the resulting neutrino mass matrix is compatible with TBM.
However, the atmospheric mass squared difference vanishes. This problem 
 is resolved, if higher-dimensional operators are included into the analysis.
In the quark sector we maintain diagonal mass matrices at leading order with the mass hierarchy of the up-quarks 
in both models. 
To generate the Cabibbo angle we consider higher-dimensional 
operators with additional $\Rep{16}_H$ fields.
Furthermore, we
calculate the mass spectrum of the heavy neutral fermions. 
Finally, the flavon potential is analyzed at the renormalizable
level.

The paper is structured as follows:
in \Secref{sec:cancellation} we describe  our GUT context and show
the prerequisites which have to be fulfilled to setup the cancellation
mechanism. In \Secref{sec:T7model}  we present the $T_7$ realization, and 
in \Secref{sec:Sigma81model} the $\Sigma(81)$
realization of this mechanism. In both setups the corrections arising from higher-dimensional 
operators with products of more than one flavon are calculated, the generation of the Cabibbo
angle is discussed as well as the mass spectrum of the heavy neutral fermions.
Additionally, the flavon potentials are presented.
We summarize our results in \Secref{sec:conclusions}.
Details of the group theory of $T_7$ and $\Sigma(81)$ as well as of 
the study of the higher-dimensional operators in both realizations 
are given in the Appendices.

\section{Cancellation of the Dirac Mass Hierarchy}
\label{sec:cancellation}

We consider an $\SO(10)$ model in which the SM fermions and the right-handed neutrinos
are unified into three $\Rep{16}_{i}$, $i=1,2,3$. 
In addition, we introduce  three  fermionic $\SO(10)$ singlets $S_{i}$.
In order to guarantee that the gauge coupling is perturbative also above the GUT scale
we assume the existence of only low-dimensional
$\SO(10)$ representations for the Higgs fields~\cite{BPWmodel,ABmodel}: $H \sim \Rep{10}$, $\Rep{16}_H$,
$\Delta \sim \overline{\Rep{16}}$ and $\Rep{45}_H$. Thus, 
no 126-dimensional scalar representation is introduced.
The non-existence of the latter causes the zeros in the mass matrix in \Eqref{neutrals}. 
The 10-dimensional representation $H$ is responsible for the Dirac masses of the fermions, 
{\it i.e.} $m_{D} \propto \langle H \rangle$. 
$\langle H \rangle$ stands for
the weak scale vacuum expectation value (VEV) for up quarks and neutrinos or for down quarks and charged leptons depending on the context.
The 16-plet scalar $\Delta$  connects the fermions in 
$\Rep{16}_{i}$ and the singlets $S_{j}$. It therefore gives rise to $m_{\nu S} \propto
\langle \Delta \rangle _{\nu}$ and $M_{NS} \propto \langle \Delta \rangle_{N}$ 
in \Eqref{neutrals}, 
where $\langle \Delta \rangle _{\nu}$ and $\langle \Delta \rangle _{N}$ 
denote the weak and the GUT scale VEVs, respectively. Note that
if there is
only one multiplet $\Delta$, the matrices $m_{\nu S}$ and $M_{NS}$ stem from the same 
coupling and are hence proportional to each other up to
renormalization group (RG) corrections. According to \Eqref{effmat} 
this leads to the proportionality $m_\nu^{LS} \propto m_{D} + m_{D}^{T}$~\cite{Barr:2003nn}.

To explain the existence of  three generations we unify $\Rep{16}_{i}$
into a $\Rep{3}$ under $G_f$. By choosing the representation
of $G_f$ to be complex we prevent the existence
of an invariant coupling $\Rep{16}_{i}\, \Rep{16}_{i}\,  H$ (if $H$ transforms trivially under $G_f$) 
which leads to
degenerate mass spectra for the fermions. Exactly for this reason the group 
$A_4$ is not applicable. In contrast to this, the transformation properties of the fermionic singlets $S_{i}$ 
are determined by the requirement to obtain a phenomenologically viable model.
It turns out that in our
two realizations it is favorable to choose the three singlets $S_i$ to transform as three
inequivalent one-dimensional representations of the flavor group $G_f$, 
instead of unifying them into a three-dimensional
representation. Obviously, a successful model should be able to describe more features of the 
fermions than just the existence of three generations. In this paper, we concentrate on models
which explain the different hierarchies of the charged fermions and the neutrinos 
through the cancellation mechanism and (some of) the prominent features of the lepton mixings, while 
simultaneously resulting in vanishing quark mixing at leading order. Nevertheless, we ensure that the Cabibbo
angle can be generated at subleading order.

The scalars in our model are separated into two groups, 
the GUT Higgs and flavon fields, in order to
disentangle the GUT and the flavor breaking scales. 
The GUT Higgs multiplets, $H$, $\Rep{16}_H$, $\Delta$ and $\Rep{45}_H$ do not transform under $G_f$,
while the flavon fields $\chi_{i}$ are gauge singlets carrying flavor indices  
\footnote{This is true, if we only consider the part relevant for neutrino masses. 
The operators required for generating the Cabibbo angle also involve GUT Higgs fields in
non-trivial one-dimensional representations of the flavor group. However, this does not alter
the statement that the GUT and the flavor symmetry breaking are disentangled in both realizations.}.
We are discussing realizations with a minimal number of flavon fields, 
which is three in our case.

As mentioned before, in order to achieve a cancellation of the mass 
hierarchy encoded in $m_{D}$
the elements of $M_{NS}$ should have a similar hierarchy.
One possibility to relate $m_{D}$ and $M_{NS}$ is to further unify
the left-handed neutrinos (and therefore $\Rep{16}_i$) and the
fermionic singlets $S_{j}$, {\it e.g.} into an $E_6$ representation~\cite{Lindner:2005pk}.
Another one is to assume that a flavor symmetry dictates the relation between $m_D$ 
and $M_{NS}$. However, when applying a flavor symmetry, in general, 
more than one Higgs field (which form non-trivial multiplets of the symmetry group) contribute
to $m_D$ and $M_{NS}$, respectively. Since these fields are in 10- and
16-dimensional representations of $\SO(10)$, respectively, it is not obvious how to properly relate
their VEVs through the Higgs potential in order to ensure that $m_D$ and $M_{NS}$ have a similar hierarchy. 
Therefore, we consider the possibility to have additional fields in 
the theory, the flavons, which are necessary to build invariants under $G_f$
\begin{equation}
\frac{\alpha}{\Lambda}\Rep{16}~ \Rep{16}~ H \chi +
\frac{\beta}{\Lambda}\Rep{16}~ S~ \Delta \chi'  +
M_{SS} S S,
\label{so10int}
\end{equation}
with either $\chi' = \chi$ or $\chi' = \chi^{\star}$. 
Here $\alpha$ and  $\beta$ are complex three-by-three matrices. 
The fact that the same flavon field (or its conjugate) enters 
both interaction terms in \Eqref{so10int} leads to the required 
correlation of the mass matrices $m_D$ and $M_{NS}$. 
The couplings involving $\Rep{16}$ in \Eqref{so10int} are non-renormalizable and suppressed 
by the cutoff scale $\Lambda$. This scale is not fixed a priori, but a natural 
choice would be the Planck scale $M_{Pl}$.
Since also the 
mass of the top quark stems from such a coupling, the ratio $\langle \chi \rangle/\Lambda$
cannot be small for all fields $\chi_i$. Thus, a careful study of
higher-dimensional operators arising from multi-flavon insertions is
mandatory \footnote{By introducing additional symmetries, such as a $\U(1)$
symmetry, one might be able to forbid all operators with more than one flavon.
However, in this paper we would like to concentrate on the simplest models with the least number of additional symmetries.}.  
We assume, for simplicity, that the
singlets $S_i$ acquire a direct Majorana mass $M_{SS}$ at the lowest order.
However, in general, this mass term is also corrected by operators involving
the flavon fields $\chi$ or could be even generated solely through these
operators. 

The idea to erase the Dirac mass hierarchy in the light
neutrino mass matrix by introducing flavon fields has been
previously addressed in~\cite{King}. There, 
 the hierarchy of  $m_D$ is
cancelled in the context of the type-I seesaw mechanism
by a quadratic hierarchy in  the Majorana mass matrix $M_{RR}$
of the right-handed neutrinos. This cancellation is complete.
Since $M_{RR}$ is strongly hierarchical, sequential dominance is
realized  which  leads (with additional constraints on the
vacuum alignment) to TBM. The gauge group is
the Pati-Salam group, and either $\SO(3)$ or
$A_4$ have been employed as $G_f$. Compared to this model our approach has the
advantage, that it can be reconciled with an embedding into $\SO(10)$
without introducing extra dimensions.

\mathversion{bold}
\section{$T_7$ Realization}
\mathversion{normal}
\label{sec:T7model}

The group $T_{7}$ is of order 21 and contains five irreducible
representations which are denoted by $\MoreRep{1}{1}$, $\MoreRep{1}{2}$,
$\MoreRep{1}{3}$ and $\Rep{3}$, $\Rep{3} ^{\star}$. The representations
$\MoreRep{1}{2}$ and $\MoreRep{1}{3}$ as well as $\Rep{3}$ and $\Rep{3} ^{\star}$
are complex conjugated to each other. $T_7$ is a subgroup of $\SU(3)$~\cite{Luhn:2 007sy}.  
This group has properties similar to those of 
the well-known group $A_{4}$ except for the crucial difference that its 
 three-dimensional representation is complex.
Due to this difference the product $\Rep{3} \times \Rep{3}$ does not
contain the invariant $\MoreRep{1}{1}$, and therefore the renormalizable coupling 
$\Rep{16}_i~ \Rep{16}_i~ H$ (for $H \sim \MoreRep{1}{1}$ under $T_7$) 
is forbidden.
It is interesting to note that $T_{7}$ is the smallest discrete group
with a complex irreducible three-dimensional representation.
In the following model 
we assume the existence of low-scale supersymmetry.

\subsection{Masses and Mixing at the Lowest Order}
\label{sec:T7lowest}

To explain the three generations of SM fermions we assign $\Rep{16}_{i}$
to $\Rep{3}$. For $H \sim \MoreRep{1}{1}$ and $\chi_{i} \sim \Rep{3}^{\star}$
the Dirac  mass matrix which results from the first term of \Eqref{so10int}
is diagonal and the VEVs of $\chi_{i}$
determine the charged fermion mass hierarchy.
In order to achieve a partial cancellation of this hierarchy in $m_{\nu}$
we assign the three fermionic  SO(10) singlets  to the three one-dimensional
representations, $S_{i} \sim \MoreRep{1}{i}$.
The Higgs multiplet $\Delta$ connecting the 16-plets $\Rep{16}_{i}$ and 
$S_{j}$ is invariant under $T_{7}$. 
Thus, also in this case we generate terms of the form as in \Eqref{so10int}.
These  assignments are collected in \Tabref{tab:T7reps}.
\begin{table}
\begin{center}
\begin{tabular}{|l||c|ccc||cc||c|}
\hline
Field & $\Rep{16}_{i}$ & $S_{1}$ & $S_{2}$ & $S_{3}$ & $H$ & $\Delta$ & $\chi_{i}$ \\\hline
\rule[0.15in]{0cm}{0cm}$\SO(10)$  & $\Rep{16}$ & $\Rep{1}$ & $\Rep{1}$ & $\Rep{1}$ &$\Rep{10}$ &$\overline{\Rep{16}}$ & $\Rep{1}$ \\
$T_7$  & $\Rep{3}$ & $\MoreRep{1}{1}$ & $\MoreRep{1}{2}$ &
$\MoreRep{1}{3}$ & $\MoreRep{1}{1}$ & $\MoreRep{1}{1}$ & $\Rep{3} ^{\star}$\\[0.04in]
\hline
\end{tabular}
\end{center}
\begin{center}
\begin{minipage}[t]{11.5cm}
\caption[]{Minimal particle content in the $T_7$ realization. $\Rep{16}_i$ and
   $S_i$ are the matter superfields, $H$ and $\Delta$ are Higgs fields and $\chi_i$ are flavons.
\label{tab:T7reps}}
\end{minipage}
\end{center}
\end{table}
The Yukawa couplings can be constructed using the Clebsch-Gordan
coefficients given in \Appref{app:T7}
\begin{equation}
\label{eq:LOT7}
\begin{split}
\mathcal{L} _{Y} &= \alpha \, ( \Rep{16}_{3} \, H \, \Rep{16}_{3} \, \chi_{1} +
\Rep{16}_{1} \, H \, \Rep{16}_{1} \, \chi_{2} + \Rep{16}_{2} \, H \, \Rep{16}_{2} \, \chi_{3} )/\Lambda
\\
& + \beta_{1} \, 
(\Rep{16}_{1} \, \chi_{1} + \Rep{16}_{2}\, \chi_{2} + \Rep{16}_{3} \, \chi_{3} )  \, \Delta \, S_{1}/\Lambda\\
& + \beta_{2} \, (\Rep{16}_{1} \, \chi_{1} + \omega
\, \Rep{16}_{2} \, \chi_{2} + \omega ^2 \, \Rep{16}_{3}  \, \chi_{3} )  \, \Delta \, S_{2}/\Lambda\\
& + \beta_{3}  \, (\Rep{16}_{1} \, \chi_{1} + \omega^2
\, \Rep{16}_{2}\, \chi_{2} + \omega \, \Rep{16}_{3} \, \chi_{3} )\, \Delta \, S_{3} /\Lambda\\
& + A \, S_{1} \, S_{1} + B \, ( S_{2} \, S_{3} + S_{3} \, S_{2} ) + \mathrm{h.c.}. 
\end{split}
\end{equation}
They  generate matrices $m_{D}$, $M_{NS}$ and $M_{SS}$ of the
form
\footnote{Throughout this work we assume that $\alpha$, $\langle H \rangle$, $\langle \Delta \rangle _N$
and $\epsilon$ are real and positive.}
\small
\begin{align}
\label{eq:matstructT7}
m_{D} &= \frac{\alpha \, \braket{H}}{\Lambda} \,
	\left( \begin{array}{ccc}
	  \braket{ \chi_{2} } & 0 & 0\\
	   0 & \braket{ \chi_{3} } & 0\\
	   0 & 0 & \braket{ \chi_{1} }
	\end{array}
	\right)\; , \\\nonumber
\begin{split}
M_{NS} &= \frac{\braket{\Delta}_N}{\Lambda} \, \left( \begin{array}{ccc}
	\beta_{1} \, \braket{ \chi_{1} } & \beta_{2} \, \braket{ \chi_{1} } & \beta_{3} \, \braket{
\chi_{1} }\\
	\beta _{1} \, \braket{ \chi_{2} } & \omega \, \beta_{2} \, \braket{ \chi_{2} } &
\omega^2 \, \beta_{3} \, \braket{ \chi_{2} }\\
	\beta_{1} \, \braket{ \chi_{3} } & \omega^2 \, \beta_{2} \, \braket{ \chi_{3} } &
\omega \, \beta _{3} \, \braket{ \chi_{3} }
	\end{array}
	\right)\\\nonumber
	&=  \frac{\braket{\Delta}_N}{\Lambda} \, \left( \begin{array}{ccc}
	\braket{ \chi_{1} } & 0 & 0\\
	0 & \braket{ \chi_{2} } & 0\\
	0 & 0 & \braket{ \chi_{3} }
	\end{array} \right) \,  \left( \begin{array}{ccc}
	1 & 1 & 1\\
	1 & \omega & \omega^2\\
	1 & \omega^2 & \omega
	\end{array} \right) \,  \left( \begin{array}{ccc}
	\beta_{1} & 0 & 0\\
	0 & \beta_{2} & 0\\
	0 & 0 & \beta_{3}
	\end{array} \right)\; ,
     \end{split}\\ \nonumber
     M_{SS} &= \left( \begin{array}{ccc}
	A & 0 & 0\\
	0 & 0 & B\\
	0 & B & 0
	\end{array}
	\right)\; ,
       \end{align}
with $\omega \equiv e^{\frac{2\pi i}{3}}$. Assuming the dominance of the DS contribution we 
obtain
\begin{equation}
m_{\nu}  \approx \left(\frac{\alpha \,
\braket{H}}{\braket{\Delta}_N} \right)^2 \, D_\chi
	\left( \begin{array}{ccc}
	\XX & \YY & \YY\\
	. & \XX & \YY\\
	. & . & \XX
	\end{array} \right) \,D_\chi,
\label{eq:mnu}
\end{equation}
where
\begin{equation}\label{eq:T7DefABtilde}
D_\chi \equiv	\diag\left(\frac{\braket{ \chi_{2} }}{\braket{ \chi_{1} }},\;
\frac{\braket{ \chi_{3} }}{\braket{ \chi_{2} 
}},\; \frac{\braket{ \chi_{1} }}{\braket{ \chi_{3} }}\right), ~~~
\tilde{A} \equiv \frac{A}{9 \beta_{1}^2}, ~~~\tilde{B} \equiv  \frac{B}{9 \beta_{2} \beta_{3}}.
\end{equation}
For simplicity, we will only consider the case in which the VEVs $\langle \chi_i \rangle$ are real
and positive. To produce the hierarchy of the up-quark masses, these VEVs have to be chosen as
\be
\frac{\braket{ \chi_{2} }}
{\braket{ \chi_{1} }} \approx \epsilon ^4, ~~~~\frac{\braket{ \chi_{3} }}
{\braket{ \chi_{1} }} \approx \epsilon ^2~~
{\rm with}~~ \epsilon \approx 0.05.
\label{eq:vev-hier}
\ee
The corresponding flavon potential is discussed 
in \Secref{sec:T7flavon}.
The ratio $\langle \chi_1 \rangle/\Lambda$ cannot be small, {\it i.e.}
\be 
\eta \equiv  \frac{\braket{ \chi_{1} }}{\Lambda} \sim \mathcal{O}(1) \; ,
\label{eq:eta}
\ee  
to guarantee the large mass of the top quark.
Using \Eqref{eq:vev-hier} we obtain from \Eqref{eq:mnu}
\begin{equation}
m_{\nu}  \approx  \left(\frac{\alpha \, \braket{H}}{\braket{\Delta}_N \, \epsilon^2}
	\right)^2
	 \, \left( \begin{array}{ccc}
	(\XX) \, \epsilon^{12} & (\YY) \, \epsilon^{6} & (\YY) \, \epsilon^{6}\\
	. & \XX & \YY \\
	. & . & \XX
	\end{array}
	\right).
\label{eq:numass}
\end{equation}
This matrix has a dominant, $\mu-\tau$ symmetric $2-3$ block,
while the elements in the first row and column are
strongly suppressed. Therefore, 
the $2-3$ mixing is maximal, $\theta_{23} = \pi/4$, 
and the two other mixing angles are very small, especially the solar mixing angle has
to be generated by additional contributions. 
The mass spectrum is normally ordered with
$m_1 \ll m_2, m_3$. For $|2\tilde A +\tilde B|<3 \, |\tilde B|$ we find
\begin{equation}\label{eq:T7masses}
m_2=\left(\frac{\alpha \,
       \braket{H}}{\braket{\Delta}_N \, \epsilon^2}\right)^2
   |2\tilde A +\tilde B|\; , \quad\quad
m_3=3\left(\frac{\alpha \, \braket{H}}{\braket{\Delta}_N \, \epsilon^2}\right)^2
   |\tilde B|
\end{equation}
and therefore
\begin{equation}
r \equiv   \frac{\Delta m_{21}^2}{\Delta m_{31}^2}=\frac{m_{2}^{2}-m_{1}^{2}}{m_{3}^{2}-m_{1}^{2}}\approx
\frac{|2\tilde A+\tilde B|^2}{9\,|\tilde B|^2}\; .
\end{equation}
The smallness of the ratio $r$ is achieved, for example, if $\tilde A \approx -\frac{1}{4} \tilde B$.
According to \Eqref{eq:numass} the DS enhances the neutrino mass terms in
the $2-3$ block, and consequently, the absolute scale of the neutrino masses  
by a factor of $\epsilon^{-4} \approx 1.6\cdot 10^5$. 
This originates from the fact that in $m_D$ and in the diagonal factor of $M_{NS}$ the VEVs of the flavons do not follow the same ordering.
For this reason, the cancellation of the
hierarchy is only partial. Furthermore, the condition 
$m_{3}  \lesssim 1 \, \eV$ implies
(for $\langle H \rangle = 174 \GeV$)
\begin{equation}
\label{eq:problemscaleT7}
\frac{(\tilde A,\tilde B)}{10^{16} {\rm GeV}} \left(\frac{10^{16} {\rm GeV}}{\braket{\Delta}_N}\right)^2 
\lesssim  2 \cdot 10^{-3} ~~~.
\end{equation}
This is not satisfied for natural values of parameters  
$\tilde A,\tilde B \sim M_{Pl} = 1.22 \cdot 10^{19} \GeV$
and $\braket{\Delta}_N \sim M_{GUT} = 2 \cdot 10^{16} \GeV$. We can fulfill
\Eqref{eq:problemscaleT7} by either lowering the scale of the 
parameters $A$ and $B$ (and thus $\tilde{A}$ and $\tilde{B}$) down to $10^{13} \GeV$ or by increasing the
scale of the VEV $\langle \Delta\rangle_{N}$ from $10^{16} \GeV$ up to $10^{19} \GeV$. 
The first possibility turns out to be favorable, since we can lower the scale 
of $A$ and $B$ by several orders of magnitude by introducing an
additional symmetry which forbids the direct mass term.
Then we have to check
whether this modification of scales alters our assumption that $m_\nu^{LS} \ll m_\nu^{DS}$. The LS term
is of the form 
\begin{equation}
m_\nu^{LS} = -2 \, \frac{\langle \Delta \rangle _\nu}{\langle \Delta \rangle _N} \, m_D
= - 2 \, \alpha \, \eta \, \frac{\langle \Delta \rangle _\nu}{\langle \Delta \rangle _N} \, \langle H \rangle \,
\left( \begin{array}{ccc}
	\epsilon^4 & 0 & 0\\
	0 & \epsilon^2 & 0\\
	0 & 0 & 1
\end{array}
\right) \; .
\label{lsls}
\end{equation}
It does not have a $1/\epsilon$ enhancement like the DS contribution and is still
of the generic size $10^{-3} \eV$. Therefore it does not change the maximal atmospheric
mixing originating from \Eqref{eq:numass}. At the same time, being diagonal, this term  does not 
generate a sizable $1-2$ mixing angle. A non-diagonal LS contribution can originate from the introduction of
a second 16-plet, $\Delta' \sim \overline{\Rep{16}}$. 
This possibility is compatible
with lowering the scale of $A$ and $B$.
This issue is discussed in \Secref{sec:T7LS}.

One might raise the question whether lowering the mass scale of the singlets  
invalidates the DS formula shown in \Eqref{eq:mnueffform}. 
This is not the case, as has been discussed in~\cite{smallMSS}. For $M_{SS} \rightarrow 0$ 
total lepton number conservation is restored and
the light neutrinos will be massless.
The right-handed neutrinos and singlets will then combine into three heavy Dirac fermions. 
We will discuss the mass spectrum of the right-handed neutrinos and the singlets in more detail
in \Secref{sec:T7_heavies}.

\subsection{Effects of Higher-Dimensional Operators}
\label{sec:T7highdim}

As the large top quark mass requires $\eta \sim \Order{1}$, 
a careful study of the higher-dimensional operators of the form
\begin{equation}
\hat{O} \left( \frac{\chi_{i}}{\Lambda}\right)^{n}
\end{equation}
$n =2,3,...$ is mandatory. 
Here $\hat{O} $ denotes 
$\Rep{16}\,\Rep{16}\, H,~ \Rep{16}\, S\, \Delta$ or $\Lambda \, S S$
\footnote{Note that there is no renormalizable coupling between the singlets $S_i$ and the flavons,
since the singlets are in one-dimensional representations of $T_7$, whereas the fields $\chi_i$
form a triplet.}.
\begin{table}
\begin{center}
\begin{tabular}{|c|c|c|}
\hline
Structure & Transformation Properties & Order in $\epsilon$\\
& under Generator $\rm A$ &\\
\hline
\rule[0.2in]{0cm}{0cm}$\chi_{1} ^{n}$ 		& $e ^{-\frac{2 \, \pi \, i}{7} \, n} \, \chi_{1} ^{n}$
& $\Order{1}$\\
$\chi_{1} ^{n-1} \, \chi_{2} $	& $e ^{- \frac{2 \, \pi \, i}{7} \,  (n+1)}
	\, \chi_{1} ^{n-1} \, \chi_{2}$
& $\Order{\epsilon^{4}}$\\
$\chi_{1} ^{n-1} \, \chi_{3} $	& $e ^{- \frac{2 \, \pi \, i}{7} \,  (n+3)}
	\, \chi_{1} ^{n-1} \, \chi_{3}$
& $\Order{\epsilon^{2}}$\\
$\chi_{1} ^{n-2} \, \chi_{2} \, \chi_{3}$ & $e ^{-\frac{2 \,
\pi \, i}{7} \, (n+4)}\, \chi_{1} ^{n-2} \, \chi_{2} \, \chi_{3}$ & $\Order{\epsilon^{6}}$\\
$\chi_{1} ^{n-2} \, \chi_{3} ^{2}$	& $e ^{-\frac{2 \, \pi \, i}{7}  \, (n-1)}
	\, \chi_{1} ^{n-2} \, \chi_{3} ^{2}$
& $\Order{\epsilon^{4}}$\\
$\chi_{1} ^{n-3} \, \chi_{3} ^{3}$	& $e ^{-\frac{2 \, \pi
\, i}{7} \, (n+2)} \, \chi_{1} ^{n-3} \, \chi_{3} ^{3}$ & $\Order{\epsilon^{6}}$\\[0.04in]
\hline
\end{tabular}
\end{center}
\begin{center}
\begin{minipage}[t]{13.5cm}
\caption[]{List of products of $\chi_{i}$ which lead to contributions down to $\Order{\epsilon^{6}}$ for
$\braket{ \chi_{1} }/\Lambda = \eta \sim \mathcal{O}(1)$,
$\braket{ \chi_{2} }/\braket{ \chi_{1} } \approx \epsilon ^4$ and
$\braket{ \chi_{3} }/\braket{ \chi_{1} } \approx \epsilon ^2$.
Note that for the order $n$ the factor $\eta^{n}$ has to be included. In the second column we show the behavior of the monomials 
under the generator $\rm A$ (see \Appref{app:T7}) which uniquely determines
their $T_{7}$ transformation properties (apart from the fact that one cannot specify
as which $T_7$ singlet the monomial transforms with this information).
\label{tab:T7_table1}}
\end{minipage}
\end{center}
\end{table}
For $\braket{ \chi_{i} }$ as given in \Eqref{eq:vev-hier}, 
the products of $\chi_{i}$ which contribute down
to $\Order{\epsilon^{6}}$  are collected in \Tabref{tab:T7_table1}.
We list all contributions up to the $6^{\rm th}$ power of $\epsilon$, $\mathcal{O}(\epsilon^6)$, 
since we have seen in the previous section that 
entries up to the $4^{\rm th}$ power, $\mathcal{O}(\epsilon^4)$, are generated in the mass matrices $m_D$ and $M_{NS}$ by the insertion of one flavon.
It should be noted that the $n^{\rm th}$ order in \Tabref{tab:T7_table1} has to be multiplied 
by $\eta^{n} <1$. Thus, contributions from such operators get more and more suppressed as $n$ increases. 
Choosing, for instance, $\eta \approx 0.47$, equivalent to $\eta\approx\epsilon^{1/4}$, 
makes it sufficient to consider operators up to $n=17$.
In \Tabref{tab:highdimopT7} presented in \Appref{app:higherdim_operators} we show as which 
component of a $T_7$ covariant the monomials given in \Tabref{tab:T7_table1} 
actually transform for arbitrary $n$.
Using these results one finds that all matrix
elements of $m_{D}$, $M_{NS}$ and $M_{SS}$ get corrections at all orders in  $\epsilon^2$,
{\it i.e.} $\Order{1}$, $\Order{\epsilon^{2}}$, \dots accompanied by appropriate suppression factors $\eta^{n}$. 
Despite the suppression factor $\eta^n$ these operators destroy the lowest order result 
for the fermion mass matrices.
In the following we will discuss how to solve this problem by adding another symmetry to the model.

\subsubsection{Majorana Masses of Singlets}

A special problem arises for 
 the masses of the singlets $S_i$. As we
have seen in the previous section, most probably the scale of the singlet mass terms  has to be (much) below 
the Planck scale to accommodate the mass scale of the light neutrinos. However, corrections stemming from 
the insertion of $n$ flavons to the singlet masses can be of the order $\Lambda \, \eta^n$ with $\Lambda$
around the Planck scale. 
Such contributions can strongly affect the absolute neutrino mass scale as well
as the lepton mixing angles, since they are in general not of the same form as the leading order structure.
In order to avoid this we invoke an additional symmetry which constrains all higher-dimensional operators and forbids the direct mass term of the singlets. In a minimal setup, no additional fields are introduced, but $M_{SS}$ is generated by operators of the structure $S\, S\, \chi^n/\Lambda^{n-1}$.  As $S_i\, S_j$ transforms as singlet, also the covariants of the type $\chi^n$ have to transform as singlets. In general, it is possible to construct all three different singlet representations at a given order $n$ in $\chi^n/\Lambda^n$ which generates all matrix elements of $M_{SS}$ at the same order. However, at the third order there is only one covariant and it transforms as the trivial singlet $\Rep{1}_1$ with respect to $T_7$. Therefore the operators of the form $S\, S\, \chi^3/\Lambda^2$ explicitly read
\begin{equation}
\label{eq:MSS_nonrenorm}
a \, S_{1} \, S_{1} \chi_{1} \chi_{2} \chi_{3}/\Lambda^2 
+ b \, (S_2 \, S_3 + S_3 \, S_2) \, \chi_1 \chi_2 \chi_3/\Lambda^2
+ \mathrm{h.c.}
\end{equation}
and lead to $M_{SS}$ as displayed in \Eqref{eq:matstructT7} with $A=a \, \eta^3 
\epsilon^6 \Lambda$ 
and $B= b \, \eta^3 \epsilon^6 \Lambda$. They give exactly the same result for $m_\nu$ 
as the matrix $M_{SS}$ stemming from a direct mass term.
However, the parameters $A$ and $B$ are 
of the order $\epsilon^6 \Lambda \approx 10^{11} \GeV$.  
So, they even overcompensate 
the factor $1/\epsilon^4$ appearing in $m_\nu$ in \Eqref{eq:numass}. To correctly adjust the 
light neutrino mass scale one also has to assume that the VEV of $\Delta$ 
is smaller than the GUT scale, $\langle\Delta \rangle_N \approx 10^{15} \GeV$
\footnote{Lowering $\langle \Delta \rangle_N$ below the GUT scale does not only 
enhance the DS contribution, but also
the LS contribution, see \Eqref{effmat}. However, this  enhancement still keeps
the LS contribution subdominant compared to the DS term.}. 

As the structure of the covariants given in \Tabref{tab:T7_table1} and \Tabref{tab:highdimopT7} for $n+7$ equals the one for $n$, we choose a $Z_7$ symmetry to suppress all higher-dimensional operators relative to the leading order. 
Then, corrections to the terms of \Eqref{eq:MSS_nonrenorm} are suppressed with a relative factor
$\eta^7$ and entries which vanish at leading order are generated 
by operators of the form $S_i S_j \chi^{10}/\Lambda^{9}$ giving contributions of 
order  $\epsilon^6 \eta^{10} \Lambda$. 
Thus, all corrections to $M_{SS}$ are well under control.  
A viable charge assignment which forbids the direct mass term of the singlets and allows operators of type $\Rep{16}\,\Rep{16}\, H\, \chi/\Lambda,~ \Rep{16}\, S\, \Delta\,\chi/\Lambda$ as well as $S\,S\,\chi^3/\Lambda^2$ is presented in \Tabref{tab:Z7}
\footnote{Obviously, the $Z_7$ symmetry can affect the form of the potential of the GUT Higgs fields. In the simplest case, we can assume
that $Z_7$ is explicitly broken in this potential.}.
\begin{table}
\begin{center}
\begin{tabular}{|l||l|lll||ll||l|}\hline
Field & $\Rep{16}_i$ & $S_1$ & $S_2$ & $S_3$ & $H$ & $\Delta$ &
$\chi_i$\\\hline
$T_7$ & $\Rep{3}$ &  $\Rep{1}_1$ &  $\Rep{1}_2$ &  $\Rep{1}_3$  &  $\Rep{1}_{1}$ &  $\Rep{1}_{1}$ &  $\Rep{3}^*$ \\
$Z_7$ & $3$ & $2$ & $2$ & $2$& $0$ & $1$& $1$\\\hline
\end{tabular}
\end{center}
\begin{center}
\begin{minipage}[t]{10.5cm}
\caption[]{\label{tab:Z7}
The $Z_7$ charge assignment of all fields. A field $\phi$ with charge $q$ transforms as 
$e^{\frac{2 \pi i}{7} q} \, \phi$ under $Z_7$.}
\end{minipage}
\end{center}
\end{table}
%

\subsubsection{Cabibbo Angle}

The additional $Z_7$ symmetry also constrains the higher-dimensional operators contributing to $m_D$ and $M_{NS}$ to operators with $7\, k+1$ flavon insertions ($k=0,1,2, ...$). 
These contributions do not spoil the leading order result.
The matrix elements which are non-vanishing at leading order are corrected by contributions
which arise in the same order of $\epsilon$ as the leading order, but with an additional suppression 
factor $\eta^7 \approx \epsilon^2$.
This is due to the above mentioned periodicity in 7 of the structure of the covariants given in \Tabref{tab:T7_table1} and \Tabref{tab:highdimopT7}.
Higher-dimensional operators lead to non-vanishing off-diagonal elements of $m_D$ which are at most of the order of $\Order{\epsilon^6\eta^8}$.
Thus we cannot generate the Cabibbo angle through them
 by simply introducing a second 10-dimensional Higgs field, $H^{\prime} \sim \Rep{10}$. 
One possibility  to obtain the Cabibbo angle is to consider
operators of the form~\cite{ABmodel,BPWmodel} 
\begin{equation}
\label{eq:thCoperatorT7}
\frac{1}{M} \left(\Rep{16}\,\Rep{16}\,\Rep{16}_H\,\Rep{16}_H^\prime \right) \, \left( \frac{\chi}{\Lambda}
\right)^n
\end{equation}
 with $M$ being the mass of the particles
mediating this interaction. Thereby, new Higgs fields $\Rep{16}_H$ and $\Rep{16}_H^\prime$ have to be
introduced. When $\Rep{16}_{H}$ acquires a 
weak scale VEV $\braket{\Rep{16}_H}_\nu$ and $\Rep{16}_H^\prime$ a GUT scale VEV
$\braket{\Rep{16}_H^\prime}_N$, the operators contribute
 to the down quark and the charged
lepton mass matrix and the size of the contributions is 
$\braket{\Rep{16}_H}_\nu\,(\braket{\Rep{16}_H^\prime}_N/M) \langle \chi \rangle^n/\Lambda^n$.
If the Higgs fields $\Rep{16}_H$, $\Rep{16}_H^\prime$ do not
transform under $T_7$ and  $Z_7$, these operators cannot lead to a sizable 
contribution to the Cabibbo angle. Therefore,  
we assume that $\Rep{16}_H$ and $\Rep{16}_H^\prime$ have non-trivial
$Z_7$ charges, $Q(\Rep{16}_H)=Q(\Rep{16}_H^\prime)=6$. 
Then the lowest dimensional operator which is invariant under $T_7$ and $Z_7$ is of the form
\begin{equation}
\label{eq:ABBPWoperator}
\frac{1}{M} \left( \Rep{16}\,\Rep{16}\,\Rep{16}_H\,\Rep{16}_H^\prime \right) \, 
\left( \frac{\chi}{\Lambda} \right)^3 \; .
\end{equation}
Its contributions to the mass matrix of the down quarks and charged leptons read
\footnote{Note that these contributions are not necessarily symmetric in flavor space. However, the
elements $(ij)$ and $(ji)$ have the same order in the parameters $\epsilon$ and $\eta$.
\label{ft1}}
\begin{equation}
\langle \Rep{16}_H \rangle _\nu \left(\frac{\langle \Rep{16}_H^\prime \rangle_N}{M}\right)\,
\left( \begin{array}{ccc}
 \mathcal{O}(\epsilon^4\eta^3) & \mathcal{O}(\eta^3) & \mathcal{O}(\epsilon^6\eta^3)\\
 . & \mathcal{O}(\epsilon^4\eta^3) & \mathcal{O}(\epsilon^2\eta^3)\\
 . & . & 0
\end{array}
\right) \; .
\end{equation}
Assuming $\langle \Rep{16} _H \rangle_\nu \approx 100 \GeV$ and 
$\langle \Rep{16}_H^\prime \rangle_N/M \approx \epsilon^2$ 
({\it e.g.} for $\langle \Rep{16}_H^\prime \rangle_N \approx 2 \cdot 10^{16} \GeV$  
the mediator scale is $M \approx 8 \cdot 10^{18} \GeV$) we find that
the $1-2$ quark mixing  angle $\vartheta_{12} \approx \eta^2 \approx 0.22$, thus reproducing 
the  Cabibbo mixing.
The observed values of the quark mixing angles $\vartheta_{13}$ and 
$\vartheta_{23}$  cannot be generated in this way, but
 by introducing operators with a different structure. 
The contributions to the diagonal elements of
the down quark and the charged lepton mass matrix are suppressed 
by at least $\epsilon^2$ (stemming
from $\langle \Rep{16}_H^\prime \rangle_N/M$) compared to $m_{D}$. 
The mass hierarchy generated in the mass matrix $m_D$ is slightly changed, {\it i.e.}  
the mass of the first 
generation is now of order $\epsilon^3 \eta$ 
instead of $\epsilon^4 \eta$. This enhancement is welcome, 
as the mass hierarchy in the down quark and
the charged lepton sector is milder than in the up-quark masses. 
However, this is still not sufficient, since the 
masses of the first two generations turn out to be too suppressed compared to the 
mass of the third generation. 
 The subleading contributions to operators of structure \Eqref{eq:ABBPWoperator}
are suppressed by a factor $\eta^7$ and carry the same suppression factors in $\epsilon$ as the 
leading order so that they can be safely neglected. Since the operators of structure
\Eqref{eq:ABBPWoperator}
also contribute to the charged lepton mass matrix, they induce a charged lepton mixing angle
of the size of the Cabibbo angle in the $1-2$ sector. This then corrects the results for the
lepton mixing angles as well. However, this contribution alone is too small to generate 
a large $1-2$ leptonic mixing. Thus, we still have to consider another mechanism 
to generate a sizable $1-2$ mixing angle in the neutrino sector. This is discussed in 
\Secref{sec:T7LS}.

\subsubsection{Light Neutrino Mass Matrix}

Calculating finally the light effective neutrino mass matrix $m_\nu$ with the corrected matrices
$m_D$, $M_{NS}$ and $M_{SS}$ we find that the first row and column of the neutrino mass matrix 
receive small corrections from higher-dimensional ($Z_7$ invariant) operators. 
The $1-2$ and $1-3$ entries are corrected by contributions of $\Order{\epsilon^4 \eta^7}$
relative to the $2-3$ sub-block.
$\nu_1$ remains approximately massless and $m_2^2$ and $m_3^2$  
receive corrections of
$\Order{\eta^7}$ relative to the leading order result.
The corrections are negligible for all mixing angles. 
Also RG corrections cannot generate a sizable $\theta_{12}$, 
since the neutrino masses have a strong 
normal hierarchy and the value of $\theta_{12}$ at leading order is small.
Therefore, we discuss in the following how a sizable $1-2$ mixing angle 
can be generated in the
neutrino sector through an additional Higgs field contributing only to the LS term.

\subsection{Contribution from the Linear Seesaw Term}
\label{sec:T7LS}

In the previous sections, the LS term has been neglected. Since the LS contribution coming from $\Delta$ alone is diagonal 
in flavor space, see \Eqref{lsls},
it cannot lead to a large solar mixing angle anyway. 
We now extend our setup by a second Higgs 16-plet which we denote as
$\Delta^\prime\sim\Rep{\overline{16}}$. In the simplest case it
has the same transformation properties under $T_7$ and $Z_7$ as the
Higgs field $\Delta$, {\it i.e.} it is a $T_7$ singlet and has a charge
$Q(\Delta^\prime)=1$ under $Z_7$.
The additional couplings are given by
\begin{equation}
\begin{split}
\mathcal{L} _{\Delta^\prime} &=  \beta^\prime_{1}  \, 
(\Rep{16}_{1} \, \chi_{1} + \Rep{16}_{2}\, \chi_{2} + \Rep{16}_{3} \, \chi_{3} )\, \Delta^\prime \, S_{1}
/\Lambda\\
& + \beta^\prime_{2}  \, (\Rep{16}_{1} \, \chi_{1} + \omega
\, \Rep{16}_{2} \, \chi_{2} + \omega ^2 \, \Rep{16}_{3}  \, \chi_{3} )\, \Delta^\prime \, S_{2}
/\Lambda\\
& + \beta^\prime_{3} \, (\Rep{16}_{1} \, \chi_{1} + \omega^2
\, \Rep{16}_{2}\, \chi_{2} + \omega \, \Rep{16}_{3} \, \chi_{3})  \, \Delta^\prime \, S_{3}/\Lambda\; .
\end{split}
\end{equation}
Note, that it is always possible to find a linear combination of $\Delta$ and
$\Delta^\prime$ with a vanishing GUT scale VEV. Therefore we assume
$\braket{\Delta^\prime}_N=0$, so that the cancellation mechanism in the
DS contribution is not affected. 
In the presence of $\Delta^\prime$ the proportionality $m_{\nu S} \propto M_{NS}$ 
is not maintained anymore and therefore non-diagonal elements in $m_\nu$ are generated. At 
the leading order the LS contribution equals 
\small
\begin{equation}
m_\nu^{LS}=-\frac{\alpha\eta\braket{H}\braket{\Delta^\prime}_\nu}{3\epsilon^2\braket{\Delta}_N}
\left(\begin{array}{ccc}
2 \left(3\frac{\braket{\Delta}_\nu}{\braket{\Delta^\prime}_\nu}+
\sum\limits_{i=1}^3 \gamma_i\right)\epsilon^6&
\sum\limits_{i=1}^3\gamma_i\omega^{1-i} &
\sum\limits_{i=1}^3\gamma_i\omega^{i-1} \\
.&  \Order{\epsilon^4} &   \Order{\epsilon^2}\\
.& . &  \Order{\epsilon^2}\\
\end{array}\right)\; ,
\end{equation}
\normalsize
where $\gamma_i \equiv \beta_i^\prime/\beta_i$ and we assume
$\braket{\Delta}_\nu\lesssim\braket{\Delta^\prime}_\nu$ such that
the main contribution is due to $\Delta^\prime$.
In order to produce a large angle $\theta_{12}$, the LS
contribution has to be comparable to the DS
contribution. The dominant terms of the neutrino mass matrix are
\small
\begin{equation}
m_\nu\approx\left(\frac{\alpha\braket{H}}{\braket{\Delta}_N\epsilon^2}\right)^2\left(\begin{array}{ccc}
-2 X
\left(3\frac{\braket{\Delta}_\nu}{\braket{\Delta^\prime}_\nu}+\sum\limits_{i=1}^{3}\gamma_i\right)\epsilon^6&
-X\sum\limits_{i=1}^3\gamma_i\omega^{1-i} &
-X\sum\limits_{i=1}^3\gamma_i\omega^{i-1} \\
. & \XX & \YY\\
. & . & \XX\\
  \end{array}\right)\; .
\end{equation}
\normalsize
The SO(10) Higgs VEVs can be adjusted such that
$X=\frac{\braket{\Delta}_N\braket{\Delta^\prime}_\nu\epsilon^2\eta}{3\alpha\braket{H}}$
leads to the correct hierarchy between the elements of the first row/column and the $2-3$
sub-block: the mass parameters of the singlets, encoded in  $\tilde A$ and $\tilde B$,  
see \Eqref{eq:T7DefABtilde}, have to be smaller than $\braket{\Delta}_N$.
The resulting mixing angles equal
\begin{equation}
\label{eq:T7LSapproxangles}
\tan 2 \theta_{12} \approx 
\frac{\sqrt{2} |(2\gamma_1-\gamma_2-\gamma_3) \, X|}{|2\tilde A+\tilde B|}\; , \;\;
\sin\theta_{13}\approx \frac{|(\gamma_2-\gamma_3) \, X|}{\sqrt{6}\,|\tilde B|} \; , \;\;
\theta_{23}\approx \frac{\pi}{4}
\end{equation}
under the assumptions $|2\tilde A +\tilde B|<3|\tilde B|$ and 
$|(\gamma_2-\gamma_3) \, X| \ll |\tilde B|$. 
Hence a large $\theta_{12}$ and small $\theta_{13}$ can be 
accommodated. 
Contributions coming from  the diagonalization of the 
charged lepton mass matrix, 
may lead to a Cabibbo angle-size contribution to $\theta_{12}$,  
together with smaller corrections to $\theta_{13}$ and $\theta_{23}$. 
For the angle  $\theta_{12}$ 
such a contribution however can be  compensated 
by an appropriate choice of the parameter combination 
$|2 \, \gamma_1 - \gamma_2 - \gamma_3|$ in \Eqref{eq:T7LSapproxangles} to match the 
experimental value. 
The light neutrino masses are also corrected by the LS contribution, 
especially $m_1$ and $m_2$ 
\begin{eqnarray}
\label{eq:LS_numasses}
&& m_1\approx\left(\frac{\alpha \, \braket{H}}{\braket{\Delta}_N \,
    \epsilon^2}\right)^2 \, |2 \, \tilde A + \tilde B| \, \left| \frac{\sin^2 \theta_{12}}{\cos 2 \, \theta_{12}} 
\right| \; , \;\;
m_2\approx \left(\frac{\alpha \, \braket{H}}{\braket{\Delta}_N \,
    \epsilon^2}\right)^2 \, |2 \, \tilde A + \tilde B| \, \left| \frac{\cos^2 \theta_{12}}{\cos 2 \, \theta_{12}} 
\right| , \; \\ \nonumber
&& m_3\approx 3 \, \left(\frac{\alpha \,
       \braket{H}}{\braket{\Delta}_N \, \epsilon^2}\right)^2  \, |\tilde B|\; .
\end{eqnarray}
As one can see, the LS term leads to large changes in the $1-2$ sector, 
but mainly preserves the $2-3$ sector. 
Thus maximal atmospheric mixing
is still a prediction of the $T_7$ realization and the other mixing angles
and masses can be fitted to the experimental data. 
For instance, from the best fit values of the 
mass squared differences and of 
$\theta_{12}$~\cite{Maltoni:2004ei} we deduce $|2 \tilde A + \tilde B|
\approx 1.13 \cdot 10^{9} \GeV$, $|\tilde B| \approx 3.38 \cdot 10^{9} \GeV$ and 
$|2 \, \gamma_1 - \gamma_2 - \gamma_3| \approx 0.0833$ for 
$X \approx 2.25 \cdot 10^{10} \GeV$ and
$\langle \Delta^\prime \rangle_\nu = 10 \GeV$.
Note that we already used here that $\langle \Delta \rangle_N \approx 10^{15} \GeV$ 
and for simplicity we set $\alpha=1$.
There is no particular prediction for the angle $\theta_{13}$ 
and it vanishes for $\gamma_2=\gamma_3$.
The neutrinos obey a normal hierarchy 
with the lightest neutrino mass being $m_1\approx 0.00424\eV$.
\Eqref{eq:LS_numasses} shows that our model allows for a non-trivial 
relation between  the ratio $m_1/m_2$
and $\theta_{12}$
\begin{equation}
\frac{m_1}{m_2} \approx \tan^2 \theta_{12}
\end{equation}
which leads to $m_1/m_2 \approx 0.437$ for the best fit value of $\theta_{12}$~\cite{Maltoni:2004ei}.
The corrections coming from the RG running below the mass scale of the lightest heavy neutral fermion are small due to the normal 
hierarchy in the light neutrino masses. 
However, the effects can be
larger above this scale. We briefly comment on these effects in the next section.
Finally, note that contributions from higher-dimensional operators 
to the LS term are also controlled by
the \Groupname{Z}{7} symmetry, as it was discussed 
in \Secref{sec:T7highdim} for the DS term.

\subsection{Masses of Right-handed Neutrinos and Singlets}
\label{sec:T7_heavies}

In this section we consider the mass spectrum of the
heavy neutral leptons, the right-handed neutrinos $N_i$   
and the singlets $S_i$. Neglecting the mixing with light neutrinos 
we can write the corresponding six-by-six mass matrix as 
\begin{equation}
\left( N, S \right) \, \left(
  \begin{array}{ccc}
    0 & M_{NS}\\
    M_{NS} ^{T} & M_{SS}
    \end{array}
\right) \, \left( \begin{array}{c} N \\ S \end{array} \right) \; .
\end{equation}
The matrices $M_{NS}$  and $M_{SS}$ are given in \Eqref{eq:matstructT7}.
As we have established in the previous subsections
to accommodate the light neutrino mass scale
and to protect the mass matrix of the singlets against 
too large contributions from higher-dimensional
operators, the VEV   $\langle \Delta \rangle _N$ 
has to be $10^{15} \GeV$
and the parameters $A$,$B$ $\approx \epsilon^6 \Lambda \approx 10^{11} \GeV$.  
We find for the mass spectrum of these heavy states (HS) 
\footnote{For this calculation we assumed that all parameters are real.}
\begin{eqnarray}
&& M_{HS \, 1} \approx \eta^2 \, 
\frac{\langle \Delta \rangle _N ^2}{|\tilde A + 2 \tilde B|} \, \epsilon^8  \;\; , \;\;\; 
M_{HS \, 2} \approx 9 \, \frac{|\beta_1 \, \beta_2 \, \beta_3|^2}{|\beta_1^2 \beta_2^2 + \beta_1^2 \beta_3^2
+ \beta_2^2 \beta_3^2|} \, |\tilde A + 2 \tilde B| \; , \\ \nonumber 
&& M_{HS \, 3,4} \approx \sqrt{3} \, \eta \, \sqrt{\frac{\beta_1^2 \beta_2^2 + \beta_1^2 \beta_3^2 + \beta_2^2 \beta_3^2}{\beta_1^2 + \beta_2^2 + \beta_3^2}} \, \langle \Delta \rangle _N \, \epsilon^2 
\;\;\; \mbox{and} \;\;\; 
M_{HS \, 5,6} \approx \eta \, \sqrt{\beta_1^2 + \beta_2^2 + \beta_3^2} \, \langle\Delta \rangle _N\; .
\end{eqnarray}
As one can see,  the four heaviest states form two pseudo-Dirac pairs, while the two lowest
lying states have a certain mass splitting. A rough estimate of the size of the masses
leads to $M_{HS \, 1} \approx 10^{8} \GeV$, $M_{HS \, 2} \approx 10^{11} \GeV$, 
$M_{HS \, 3,4} \approx 10^{12} \GeV$ and
$M_{HS \, 5,6} \approx 10^{15} \GeV$. For our numerical example which 
we discuss at the end of \Secref{sec:T7LS}
we can choose the following values of $\tilde A$ and $\beta_i$ to arrive at the numbers given above
\footnote{The choice of the parameters $\beta_i$ is more or less free 
(they should remain in the perturbative regime), 
since we can always use the couplings $\beta_i^{\prime}$ to appropriately adjust the values of $\gamma_i$ 
which play a role in the calculation of the mixing angles above. For $\tilde B$ we simply took 
$\tilde B \approx 3.38 \cdot 10^{9} \GeV$.}
\begin{equation}
\tilde A \approx \; -1.12 \cdot 10^{9} \GeV , \;\; 
\beta_1 \approx 3.34 \; , \;\; 
\beta_2 \approx 1.34 \;\;\; \mbox{and} \;\;\; 
\beta_3 \approx 2.15 \; .
\end{equation}
The masses of the heavy states are then $M_{HS \, 1} \approx 1.49 \cdot 10^{9} \GeV$, 
$M_{HS \, 2} \approx 6.06 \cdot 10^{10} \GeV$, 
$M_{HS \, 3,4} \approx 4.35 \cdot 10^{12} \GeV$ and
$M_{HS \, 5,6} \approx 1.97 \cdot 10^{15} \GeV$.
Analyzing the decomposition of the mass eigenstates we find that the state with the smallest mass
$M_{HS \, 1}$ mainly consists of the second right-handed neutrino $N_2$, the one with $M_{HS \, 2}$
is a mixture of the three singlets, the states with $M_{HS \, 
3,4}$ decompose into the third
right-handed neutrino and the singlets, while the states with the largest masses mainly involve
the first right-handed neutrino $N_1$ and the three singlets.
Due to this fact we expect that
 above $M_{HS2}$ larger RG corrections to the light neutrino masses and mixings
originate from the  $\Order{1}$ couplings of the singlets to the first 
left-handed neutrino and above $M_{HS 3,4}$ 
additionally from the $\Order{1}$ coupling of the third right-handed neutrino 
to the left-handed neutrinos.

\subsection{Flavon Superpotential}
\label{sec:T7flavon}

Here we briefly comment on the flavon superpotential and on a possibility to achieve the 
vacuum structure shown in \Eqref{eq:vev-hier}.  
The potential can be minimized in the supersymmetric limit, since effects from soft 
breaking terms are expected to be negligible. The renormalizable part of the superpotential 
without the $Z_7$ symmetry, which is necessary to control the higher order corrections to the 
fermion  mass matrices, has the form
\begin{equation}\label{eq:T7flavon}
W_{\chi} = \kappa \, \chi_{1} \, \chi_{2} \, \chi_{3} 
\end{equation}
where $\kappa$ is a dimensionless coupling constant. 
As one can see, the F-terms, $\kappa \, \chi_{i} \, \chi_{j}$, $i<j$, $i,j \in \{1,2,3\}$ are all
zero, if (at least) two of the three VEVs $\langle \chi_i \rangle$ vanish. Therefore, we can
choose the minimum in which only $\langle \chi_1 \rangle$ is non-vanishing which is a good lowest
order approximation to the vacuum structure shown in \Eqref{eq:vev-hier}. Note that the value of
$\langle \chi_{1} \rangle$ is not fixed by the potential.

If we impose the $Z_7$ symmetry on the flavon potential, 
the term in \Eqref{eq:T7flavon} is forbidden. 
The lowest order terms of the flavon potential are then
\[
W_{\chi}= \frac{a_{1}}{\Lambda^{4}} \, (\chi_{1} ^{7} + \chi_{2} ^{7} + \chi_{3}  ^{7}) 
+ \frac{a_{2}}{\Lambda^{4}} \, (\chi_1 \, \chi_2 \, \chi_3) \, 
(\chi_{1} ^{3} \, \chi_{3} + \chi_{1}\, \chi_{2}^3 + \chi_{2} \, \chi_{3} ^{3})\; .
\]
However, the F-terms derived from this superpotential 
have no configuration $\braket{ \chi_{1} } \neq 0$ and $\braket{ \chi_{2,3} } =0$
as solution. Thus, \Groupname{Z}{7} should be explicitly broken in the flavon superpotential
or other fields should exist apart from the flavons $\chi_{i}$.
One possibility to reconcile the VEV structure 
and the \Groupname{Z}{7} symmetry is to introduce a new
 field $\varphi$ with  $T_7$ and $Z_7$ properties $\left(\Rep{3}^{*},5\right)$
\footnote{Here we additionally have to assume the existence of a $\U(1)_R$ symmetry under
which the fermions of the model transform with charge $+1$, the Higgs fields and flavons acquiring 
VEVs with $0$ and the field $\varphi$ has charge $+2$. The flavon superpotential then has to be
linear in the field $\varphi$. This is done along the lines of~\cite{AF2}. 
In the simplest case we can assume that the $\U(1)_R$ is explicitly broken in the GUT Higgs potential.}.  
The renormalizable flavon superpotential, 
\begin{equation}
 W_{\chi}= \kappa\left( \varphi_1\chi_2\chi_3 + \varphi_2\chi_1\chi_3 + \varphi_3\chi_1\chi_2\right) \; ,
\end{equation}
allows the configuration  $\braket{ \chi_{1} } \neq 0$ and $\braket{ \chi_{2,3} } =0$.
In order to study the issue whether the hierarchical structure of the VEVs $\langle \chi_i \rangle$
can be achieved with such a potential we have to discuss 
the contributions to the potential 
arising from higher-dimensional operators consisting of more than two flavons and the field $\varphi$. 
This can be done in a similar fashion as done for the contributions to the fermion mass matrices.
However, such a study is beyond the scope of this paper. Also the fact that terms involving flavon fields might  
arise as corrections to the GUT Higgs potential is not discussed here.
 
\mathversion{bold}
\section{$\Sigma(81)$ Realization}
\label{sec:Sigma81model}
\mathversion{normal}

The $\Sigma(81)$ realization of the cancellation mechanism differs from the $T_7$ one in 
two aspects: $(i)$  
the cancellation of the up-quark mass hierarchy in $m_{\nu}$ is complete 
and $(ii)$ it requires a non-supersymmetric framework, since the fields $\chi_{i}$
are involved in the coupling of $\Rep{16}_{i}$ to $S_{j}$, whereas
their complex conjugates $\chi_{i}^{\star}$ appear
in the Yukawa couplings $\Rep{16}_{i} \, \Rep{16}_{j} H$.
The group $\Sigma(81)$, previously discussed in~\cite{Ma:2006ht,Ma:2007ku},  
has one- and three-dimensional
representations. Its order is 81 and therefore it has nine one-dimensional,
$\MoreRep{1}{i}$, $\rm i=1,...,9$,  and eight three-dimensional
$\MoreRep{3}{i}$, $\rm i=1,...,8$, representations.   
Apart from $\MoreRep{1}{1}$ all of them
are complex. Contrary to $T_7$, $\Sigma(81)$ is a subgroup of $\U(3)$, but not of $\SU(3)$.
Further group theoretical aspects of  $\Sigma(81)$ are summarized in \Appref{app:Sigma81}.

\subsection{Masses and Mixing at the Lowest Order}
\label{sec:Sigma81lowest}

The three generations of fermions $\Rep{16}_{i}$ are
assigned to one of the six faithful \footnote{A faithful representation of a group
has as many distinct representation matrices as elements exist in the group.} three-dimensional
representations, $\MoreRep{3}{i}$, $\rm i=1,...,6$. Without loss
of generality we choose $\MoreRep{3}{1}$. To arrive at
a diagonal $m_{D}$ we assign  $H$ to $\MoreRep{1}{1}$ and
$\chi_{i}$ to $\MoreRep{3}{2}$,  so that $\chi_{i}^{\star}$
contributes to $\left(m_{D}\right)_{ii}$ only. The SO(10) singlets, $S_{i}$, 
transform as  $\MoreRep{1}{1}$, $\MoreRep{1}{2}$ and $\MoreRep{1}{3}$ and the Higgs field
$\Delta$ as $\MoreRep{1}{1}$. 
These properties are summarized in  \Tabref{tab:Sigma81reps}.
\begin{table}
\begin{center}
\begin{tabular}{|l||c|ccc||cc||c|}
\hline
Field & $\Rep{16}_{i}$ & $S_{1}$ & $S_{2}$ & $S_{3}$ & $H$ & $\Delta$ & $\chi_{i}$ \\\hline
\rule[0.15in]{0cm}{0cm}$\SO(10)$  &  $\Rep{16}$ &  $\Rep{1}$ & $\Rep{1}$ &
$\Rep{1}$ &$\Rep{10}$ &$\overline{\Rep{16}}$ &$\Rep{1}$ \\
$\Sigma(81)$  & $\MoreRep{3}{1}$ & $\MoreRep{1}{1}$&
$\MoreRep{1}{2}$&$\MoreRep{1}{3}$& $\MoreRep{1}{1}$& $\MoreRep{1}{1}$ & $\MoreRep{3}{2}$\\[0.04in]
\hline
\end{tabular}
\end{center}
\begin{center}
\begin{minipage}[t]{11cm}
\caption[]{Particle assignment in the $\Sigma(81)$ realization. $\Rep{16}_i$ and
   $S_i$ are fermions, $H$ and $\Delta$ are Higgs fields and $\chi_i$ are
   flavons. Note that $\MoreRep{3}{2}$ is equivalent to $\MoreRep{3}{1}^*$.\label{tab:Sigma81reps}}
\end{minipage}
\end{center}
\end{table}
As a result the matrix structure of $M_{NS}$ and $M_{SS}$
is the same as in the case of $T_7$, see \Eqref{eq:matstructT7}, while the matrix $m_{D}$ is of the form
\begin{equation}\label{eq:mDSigma81}
m_{D} = \frac{\alpha \, \braket{H}}{\Lambda} \,
	\left( \begin{array}{ccc}
	  \braket{ \chi_{1} } ^{\star} & 0 & 0\\
	   0 & \braket{ \chi_{2} } ^{\star} & 0\\
	   0 & 0 & \braket{ \chi_{3} } ^{\star}
	\end{array}
	\right) \; .
       \end{equation}
As in the $T_7$ realization we assume that the VEVs $\langle \chi_i \rangle$ 
can be chosen as real and positive.
To reproduce  the up-quark masses we take 
\be
\frac{\braket{ \chi_{1} }}
{\braket{ \chi_{3} }} \approx \epsilon ^4, ~~~~\frac{\braket{ \chi_{2} }}
{\braket{ \chi_{3} }} \approx \epsilon ^2~~
{\rm and}~~ \eta \equiv\frac{\braket{\chi_3}}{\Lambda}\sim \mathcal{O}(1) 
~~{\rm with}
~~\epsilon \approx 0.05 \; .
\label{eq:S81vev-hier}
\ee
We assume again the dominance of the DS contribution in the 
light neutrino mass matrix, which  is given by
\begin{equation}
\label{eq:sigma81_numass_2}
m_{\nu} \approx 
\left(\frac{\alpha \, \braket{H}}{\braket{\Delta}_N} \right)^2 \,
	\left( \begin{array}{ccc}
	\XX & \YY & \YY\\
	. & \XX & \YY\\
	. & . & \XX
	\end{array} \right)
\end{equation}
with $\tilde A$ and $\tilde B$ being defined in \Eqref{eq:T7DefABtilde}. 
The hierarchy of the up-quark masses encoded in the VEVs of $\chi_{i}$
is completely erased in $m_{\nu}$ without any further assumptions on
$\braket{\chi_{i}}$ and/or the couplings.
Thus, $\langle \Delta \rangle_N$ and $A$ and $B$ ($\tilde A$ and $\tilde B$)
are of their generic size, $\langle \Delta \rangle_N = 2 \cdot 10^{16} \GeV$ and $A,B \sim M_{Pl} = 1.22 \cdot 10^{19} \GeV$,
to produce a light neutrino mass scale of $1 \eV$. As always $\langle H \rangle$ is fixed to $174 \GeV$.
Note that the form of $m_\nu$ in \Eqref{eq:sigma81_numass_2}
is the most general one invariant under  $S_3$ 
~\cite{Chen:2007zj} permuting the three neutrino generations.
The matrix $m_{\nu}$ is diagonalized by the TBM matrix, however the mass spectrum   
contains two equal masses  
\be
\label{eq:Sigma81_numasses}
m_2=3 \left(\frac{\alpha\braket{H}}{\braket{\Delta}_N}\right)^2  |\tilde A| 
\;\;\; \mbox{and} \;\;\;
m_1= m_3=3\left(\frac{\alpha\braket{H}}{\braket{\Delta}_N}\right)^2 |\tilde B| \; .
\ee
Here $m_2$ corresponds to the state with tri-maximal mixing and  
$m_3$ is the mass of the state with bimaximal mixing which is degenerate with the state
with mass $m_1$. Thus, the atmospheric mass squared difference $\Delta m_\mathrm{31}^2$ vanishes.
The solar mass squared difference is given by
\be\label{eq:Sigma81dmsdma}
\Delta m_\mathrm{21}^2=
9 \left(\frac{\alpha\braket{H}}{\braket{\Delta}_N}\right)^4  (|\tilde
   A|^2-|\tilde B|^2)\; .
\ee
As one can see, we have to assume that $|\tilde A| \approx |\tilde B|$
for $\Delta m_{21} ^2$ being small. This corresponds to a quasi-degenerate neutrino mass
spectrum~\cite{Chen:2007zj}. $\Delta m_{31}^2$ can be generated
by higher-dimensional corrections as will be discussed in the next
section.

The Majorana mass matrix $M_{NN}$ of the right-handed neutrinos can be obtained by the seesaw formula due to the hierarchy  $M_{NS} \ll  M_{SS}$
\begin{equation}
\label{eq:seesaw_MNN}
M_{NN} \approx - M_{NS} \, M_{SS}^{-1} \, M_{NS}^{T}. 
\end{equation}
Then using the matrices $M_{NS}$ and $M_{SS}$ of \Eqref{eq:matstructT7} we obtain  
the analytic formulae for  the right-handed 
neutrino masses (for real parameters)
\begin{equation}
M_{NN \, 1} \approx \epsilon^8 \, \frac{\langle \Delta \rangle_N ^2 \eta^2}{|\tilde A + 2 \tilde B|} 
\; ,\;\;
M_{NN \,2} \approx \epsilon^4 \, \langle \Delta \rangle_N ^2 \eta^2 \, \left|\frac{\tilde A + 2 \tilde B}{3 \, \tilde B \, (2 \tilde A + \tilde B)} \right| 
\; ,\;\;
M_{NN \, 3} \approx \langle \Delta \rangle_N ^2 \eta^2 \, \left|\frac{2 \tilde A + \tilde B}{9 \, \tilde A \, \tilde B} \right|
\end{equation}
with $\tilde{A}$ and $\tilde{B}$ according to \Eqref{eq:T7DefABtilde}. Hence, the masses of the right-handed neutrinos are expected to be strongly hierarchical
\footnote{However, we would like to mention
that the given estimate of the masses of the right-handed neutrinos is only a rough one, since the actual
masses in a numerical example can easily differ by more than one order of magnitude (see 
\Secref{sec:S81neutrals}).}
\be 
\{\epsilon^8, \epsilon^4, 1 \} \cdot \eta^2 \cdot \left( \langle \Delta \rangle_N ^2/ 
M_{Pl}\right)
\approx \{ 283 \GeV, 4.53 \cdot 10^7 \GeV, 7.24 \cdot 10^{12} \GeV\} 
\ee 
for $\eta \approx 0.47$. This strong hierarchy is due to the  hierarchy 
in $M_{NS}$ and we find 
one very low lying state in the spectrum of the right-handed neutrinos. 
Its phenomenology, detection and observable consequences are determined by the mixing with active neutrinos.
The mixing angle with the flavor state $\nu_f$ is described by
\begin{equation}
\tan\theta_f \approx \frac{(m_D U_{NN})_{f1}}{M_{NN \, 1}}
\end{equation}
where $U_{NN}$ is the mixing matrix which diagonalizes the Majorana
mass matrix of the right-handed neutrinos \Eqref{eq:seesaw_MNN},
$M_{NN} = U_{NN}^* M_{NN}^{\mathrm{diag}} U_{NN}^\dagger$.
If there is no strong hierarchy of the elements of $m_\nu$, {\it i.e.} $\tilde{A} + 2 \tilde{B}$ and
$\tilde{A} - \tilde{B}$ are of the same order,
an estimate of the angle is straightforward
\begin{equation}
\tan\theta_f \approx \frac{\alpha \braket{H} |\tilde A+2\tilde B|}{\braket{\Delta}_N^2\eta\,\epsilon^4}\sim \frac{\braket{H} M_{Pl}}{\braket{\Delta}_N^2\eta\,\epsilon^4}\approx  2 \cdot 10^{-6} \; .
\end{equation}
It is much below the sensitivity of present and future
experiments which is $\sim 10^{-2} - 10^{-1}$~\cite{RHNeutrinoAtCollider}. In particular, the rate
of production of $N_1$ at LHC is negligible.

The masses of the singlets are to a good approximation given by
$|A|$, $|B|$ and $|B|$,  so that two of them are nearly degenerate.

\subsection{Effects of Higher-Dimensional Operators}
\label{sec:S81highdim}

The number of the higher-dimensional operators containing more than one
flavon $\chi_{i}$ 
increases in comparison to the $T_7$ realization, 
since in a non-supersymmetric theory also the complex conjugated fields 
$\chi_{i} ^{\star}$ have to be taken into account.
The general structure of the operators in the Lagrangian is 
\begin{equation}
\hat{O} \frac{\chi_{i} ^{n_{1}} \, (\chi_{j}^{\star}) ^{n_{2}}}
{\Lambda^{n}}  
\end{equation}
with $n_{1} + n_{2} \equiv  n \geq 2$, $n_{1,2}=0,1,2,...$. 
Again, $\hat{O}$ denotes $\Rep{16}\,\Rep{16}\, H,~ \Rep{16}\, S\,\Delta$ or  $\Lambda \, S S$. 
To study the contributions from these higher-dimensional operators 
we determine the leading order terms in $\epsilon^2$ up to
$\Order{\epsilon^{6}}$ for the VEVs of the form as given in
\Eqref{eq:S81vev-hier}. This is done analogously  to the $T_7$ case.
As above, we assume $\eta \approx \epsilon^{1/4} \approx 0.47$.
The relevant monomials in the fields 
$\chi_{i}$ and $\chi_{i} ^{\star}$ are displayed in \Tabref{tab:Sigma81_table1}.
Notice that the number of relevant operators increases with the number of flavons $n$.
This could compensate the suppression factor $\eta^n$ which arises for any operator
containing $n$ flavon fields.
\begin{table}[t]
\begin{center}
\begin{tabular}{|c|c|c|}
\hline
Order in $\epsilon$ & Operator Structure & No. of Operators\\
& & at Order $n$\\
\hline
\rule[0.15in]{0cm}{0cm}$\Order{1}$ & $\chi_{3} ^{m} \, (\chi_{3} ^{\star}) ^{n-m} \,\, (m=0,...,n)$& $n+1$\\[0.04in]
\hline
$\Order{\epsilon^{4}}$ &
$\chi_{3} ^{m} \, (\chi_{3} ^{\star})^{n-1-m} \, \chi_{1}^{(\star)}
\,\, (m=0,...,n-1)$ & $2 \, n$\\[0.04in]
\hline
$\Order{\epsilon^{2}}$ &
$\chi_{3} ^{m} \, (\chi_{3} ^{\star})^{n-1-m} \, \chi_{2}^{(\star)}
\,\, (m=0,...,n-1)$ & $2 \, n$\\[0.04in]
\hline
$\Order{\epsilon^{6}}$ &
$\chi_{3} ^{m} \, (\chi_{3} ^{\star})^{n-2-m} \, \chi_{1} \, \chi_{2}
\,\, (m=0,...,n-2)$ & $4 \, (n-1)$\\[0.04in]
  & $\chi_{3} ^{m} \, (\chi_{3} ^{\star})^{n-2-m} \, \chi_{1}^{\star} \, \chi_{2} \,\, (m=0,...,n-2)$&\\[0.04in]
& $\chi_{3} ^{m} \, (\chi_{3} ^{\star})^{n-2-m} \, \chi_{1} \, \chi_{2} ^{\star} \,\, (m=0,...,n-2)$&\\[0.04in]
& $\chi_{3} ^{m} \, (\chi_{3} ^{\star})^{n-2-m} \, \chi_{1}^{\star} \,
\chi_{2}^{\star} \,\, (m=0,...,n-2)$&\\[0.04in]
\hline
$\Order{\epsilon^{4}}$ &
$\chi_{3} ^{m} \, (\chi_{3}^{\star})^{n-2-m} \, \chi_{2} ^{2} \,\,
(m=0,...,n-2)$ & $3 \, (n-1)$\\[0.04in]
& $\chi_{3} ^{m} \, (\chi_{3}^{\star})^{n-2-m} \, \chi_{2} \, \chi_{2}^{\star}
\,\, (m=0,...,n-2)$&\\[0.04in]
& $\chi_{3} ^{m} \, (\chi_{3}^{\star})^{n-2-m} \, (\chi_{2}^{\star}) ^{2} \,\,
(m=0,...,n-2)$&\\[0.04in]
\hline
$\Order{\epsilon^{6}}$ &
$\chi_{3} ^{m} \, (\chi_{3} ^{\star})^{n-3-m} \, \chi_{2} ^{3} \,\,
(m=0,...,n-3)$ & $4 \, (n-2)$\\[0.04in]
& $\chi_{3} ^{m} \, (\chi_{3} ^{\star})^{n-3-m} \, \chi_{2} ^{2}
\, \chi_{2} ^{\star} \,\,
(m=0,...,n-3)$&\\[0.04in]
& $\chi_{3} ^{m} \, (\chi_{3} ^{\star})^{n-3-m} \, \chi_{2} \,
(\chi_{2} ^{\star})^{2} \,\,
(m=0,...,n-3)$&\\[0.04in]
& $\chi_{3} ^{m} \, (\chi_{3} ^{\star})^{n-3-m} \, (\chi_{2} ^{\star})^{3} \,\,
(m=0,...,n-3)$ &\\[0.04in]
\hline
\end{tabular}
\end{center}
\begin{center}
\begin{minipage}[t]{16cm}
\caption[]{List of products of $\chi_{i}$ and $\chi_{j} ^{\star}$ which lead to contributions down
to $\Order{\epsilon^{6}}$ to the entries of the mass matrices $m_D$, $M_{NS}$ and $M_{SS}$
under the assumption that 
$\braket{\chi_{1}}/\braket{\chi_{3}} \approx \epsilon ^4$,
$\braket{\chi_{2}}/\braket{\chi_{3}} \approx \epsilon ^2$ and
$\braket{\chi_3}/\Lambda = \eta \sim \mathcal{O}(1)$.
In the third column we list the number of operators with a certain structure. 
As one can see, this number depends linearly on $n$.
Note that also here a monomial with $n$ fields leads to an additional 
suppression factor $\eta^{n}$.
\label{tab:Sigma81_table1}}
\end{minipage}
\end{center}
\end{table}
The transformation properties of the monomials under $\Sigma(81)$
are  given  in \Tabref{tab:highdimopSigma81} of \Appref{app:higherdim_operators}.
Using these properties we find that for the elements of the matrices $m_{D}$, 
$M_{NS}$  and $M_{SS}$,  which do not vanish at the leading order,
the corrections are of the same order in $\epsilon^2$ as the leading contribution, 
and/or have a higher power of $\epsilon^2$. Therefore, the hierarchy of the elements 
in the small parameter $\epsilon$ is not destroyed by the corrections. 
These elements will be of the form 
$(m_{D})_{ij} = (m_{D})_{ij}^{\mathrm{LO}} \, (1 + \mathcal{O}(\eta^{k}))$ and 
$(M_{NS})_{ij} = (M_{NS})_{ij}^{\mathrm{LO}} \, (1 + \mathcal{O}(\eta^{l}))$ 
with $k,l \geq 1$, respectively, if the corrections are taken into account.
 The off-diagonal elements of $m_{D}$ no longer vanish and the full matrix 
has  the form
\be\label{eq:Sigma81HDOpmD}
m_D
\sim  \alpha \, \braket{H}   \left(\begin{array}{ccc}
\epsilon^4 \eta(1 +    \Order{\eta^2}) & \Order{\epsilon^6\eta^2} &
\Order{\epsilon^4\eta^2}\\
. &  \epsilon^2 \eta(1 + \Order{\eta^2}) & \Order{\epsilon^2\eta^2} \\
. & . & \eta (1 + \Order{\eta})\\
\end{array}
\right)\; .
\ee
As one can see, the off-diagonal elements are smaller than the corresponding 
(in the same row or column) diagonal elements, at least by a factor $\eta$. 
Consequently, the hierarchy of the eigenvalues of $m_D$ is not changed by these elements. 
Corrections to the diagonal elements,  
which are of the same order as the leading order term, will 
require a re-adjustment of the parameters in a numerical fit to the experimental data. 
The contributions to the Majorana mass matrix $M_{SS}$ of the singlets arise from
insertions of at least two flavons, since the singlets $S_i$ transform as one-dimensional
representations under $\Sigma(81)$, whereas the flavons form a triplet. At the level
of two flavon insertions all elements of $M_{SS}$ receive contributions of the order
$\eta^2 \Lambda$. Assuming that $A$ and $B$ as well as the 
cutoff scale
$\Lambda$ are of order $M_{Pl}$,  we conclude that the corrections to $M_{SS}$ can amount to
$\eta^2 \approx 0.22$ of the leading order terms. Since the matrix structure of 
these subleading corrections  differs from the one of the leading order, 
these corrections could be important.
 
\subsubsection{Cabibbo Angle}

The simplest possibility to generate the quark mixings would 
be to assume the existence of a second Higgs 
$H^\prime$ transforming as $\Rep{10}$ under $\SO(10)$ and trivially under $\Sigma(81)$. 
Considering this possibility we see from \Eqref{eq:Sigma81HDOpmD} 
that the induced mixing angles are of order
\be
\label{eq:Sigma81HDquarkmixings}
\left(\vartheta_{12},\,\vartheta_{13},\,\vartheta_{23}\right) 
\sim\left(\order{\epsilon^4\eta},\,\order{\epsilon^4\eta},\,\order{\epsilon^2\eta}\right)\; .
\ee
These are, however, much smaller than the 
observed mixing angles. Similarly to the $T_7$ realization, 
it is possible to generate the Cabibbo angle 
by introducing operators of the form
\begin{equation} 
\frac{1}{M} \, \left(\Rep{16}\,\Rep{16}\,\Rep{16}_H\,\Rep{16}_H^\prime \right) \, 
\frac{\chi^n (\chi^\star)^m}{\Lambda^{n+m}}
\label{16prime}
\end{equation}
with at least one of the Higgs 16-plets transforming as a non-trivial one-dimensional representation 
under $\Sigma(81)$, $\MoreRep{1}{i}$, $\rm i=2,...,9$.
For instance, we can assume that $\Rep{16}_H$ (which obtains a weak scale VEV) 
transforms as $\MoreRep{1}{4}$, 
while $\Rep{16}_H^\prime$ (having a GUT scale VEV) is invariant. 
Through this we ensure that these Higgs 
fields only break $\Sigma(81)$ at the weak and not 
at the GUT scale. Therefore  the flavor and the GUT symmetry breaking are still mainly induced by different
sets of fields.
The operators of structure \Eqref{16prime} 
yield contributions to the down quark and 
the charged lepton mass matrix of the form
\footnote{See footnote \ref{ft1}.}
\begin{equation}
\label{eq:ABBPWoperator_Sigma81}
\langle \Rep{16}_H \rangle _\nu \left(\frac{\langle \Rep{16}_H^\prime \rangle_N}{M}\right)\,
\left( \begin{array}{ccc}
 \mathcal{O}(\epsilon^6 \eta^3) & \mathcal{O}(\eta) & \mathcal{O}(\epsilon^2 \eta)\\
 . & \mathcal{O}(\epsilon^6 \eta^3) & \mathcal{O}(\epsilon^4 \eta)\\
 . & . & \mathcal{O}(\epsilon^6 \eta^3)
\end{array}
\right) \; .
\end{equation}
The contributions to the diagonal elements are in general very small and therefore always
subleading. Using \Eqref{eq:Sigma81HDOpmD} and \Eqref{eq:ABBPWoperator_Sigma81}
we find that a correct value of the $1-2$ mixing, $\epsilon^{1/2} \approx 0.22$, can be obtained 
if $\langle \Rep{16}_H \rangle_\nu \approx 100 \GeV$ and   
$\langle \Rep{16}_H^\prime \rangle_N/M \approx \epsilon ^{5/2}$. 
The latter indicates that the mediator scale $M$ is of the order of the Planck scale. 
The mass hierarchy of the charged leptons 
and down quarks is expected to be $\{ \epsilon^3 \eta, \epsilon^2 \eta, \eta\}$. 
This is similar to the $T_7$ case.
The corrections to the matrix elements shown in 
\Eqref{eq:ABBPWoperator_Sigma81} due to multi-flavon insertions
are at most of the same order in $\epsilon$ as the leading contribution.  
In addition they are suppressed by some higher
power of $\eta$. This is the same as for the corrections to the matrices
$m_D$ and $M_{NS}$.
As in the  $T_7$ model, the $1-3$ and $2-3$ quark mixing angles are  too 
small. However, we can
expect that they are generated by operators with a structure 
different from the ones discussed
here. Again, 
we expect that the contributions from the operators of structure \Eqref{16prime}
 to the charged lepton mass matrix
result in a Cabibbo angle-size mixing in the
$1-2$ sector, which can have a certain impact on the lepton mixings.

\subsubsection{Light Neutrino Mass Matrix}

Finally, we discuss the corrections 
 to the effective light neutrino mass matrix coming from higher-dimensional operators. 
There are no simple analytic formulae for neutrino
masses and leptonic mixing angles, because the elements of $m_\nu$ are all of the same 
order, if there are
no further restrictions on the couplings. We only indicate some features 
which can be directly deduced from the general expression of the mass matrix 
and show with a numerical example
that a viable neutrino mass matrix can
be obtained.
One can see that $(i)$ $\Delta m_\mathrm{31}^2$ is proportional to  $\eta$ 
and it does not depend on corrections coming from $M_{SS}$ at leading order,
since these are proportional to $\eta^2$ and
$(ii)$ contributions of higher-dimensional operators yield differences 
in the diagonal elements of the mass matrix \Eqref{eq:sigma81_numass_2}
which are of relative order $\Order{\eta}$. 
Therefore, to reproduce a large atmospheric mixing angle one should require $|\tilde A -\tilde B| >
\eta |\tilde A| , \eta |\tilde B|$. Taking into account the fact that
the smallness of the solar mass squared difference
$\Delta m_{21}^2$
enforces $|\tilde A|\approx|\tilde B|$, as explained above, we find that
the relative phase of $\tilde A$ and $\tilde B$ has to be 
around $\pi$ such that $\tilde B \approx - \tilde A$.
This is used as a restriction in our numerical search whose results are detailed in the
next section.

\subsection{Phenomenology of Neutral Fermions}
\label{sec:S81neutrals}

We perform a numerical search to show that through
the inclusion of higher-dimensional operators the model can accommodate neutrino masses
and lepton mixings. In our search we consider a certain (small) set of these operators 
which contribute to $m_D$ and $M_{NS}$. Their coefficients are real random numbers 
whose absolute value lies in the interval
$[0.1,\eta^{-1}]\approx [0.1, 2.1]$. One example is
\footnote{The up-quark masses $m_u$ and $m_c$ are a factor of
two too small at the GUT scale for $m_t(M_{GUT})$ of about $79 \GeV$~\cite{Das:2000uk}. 
Contributions either from certain multi-flavon insertions not taken into
account here or from operators with an $\SO(10)$ structure different from $\Rep{16} \, \Rep{16} \, H$
may lift the masses of the up and the charm quark.} 
\begin{subequations}
\begin{align}
m_D&=\left(\begin{array}{ccc}
1.74664 \cdot 10^{-6}&0& 7.38353 \cdot 10^{-7}\\
. & 6.99757\cdot 10^{-4}& 2.98637 \cdot 10^{-5}\\
.&.&0.454331\\
\end{array}\right)\braket{H}\;,\\
M_{NS}&=\left(\begin{array}{ccc}
5.57371 \cdot 10^{-6} & 4.80734 \cdot 10^{-6} & 4.00556 \cdot 10^{-6}\\
2.23041 \cdot 10^{-3} & 1.92369 \cdot 10^{-3} \,\omega & 1.60262 \cdot 10^{-3}\,\omega^2\\
1.03422 & 0.886329 \,\omega^2& 0.702275 \,\omega\\
\end{array}\right) \braket{\Delta}_N\; ,\\
M_{SS}&=\left(\begin{array}{ccc}
0.884095 & 0 &  0 \\
. & 0 & -1\\
. & . & 0 \\
\end{array}\right) M_{Pl}\; .
\end{align}
\end{subequations}
Using $\langle H \rangle = 174 \GeV$, $\langle \Delta \rangle_N = 2 \cdot 10^{16} \GeV$ and
$M_{Pl} = 1.22 \cdot 10^{19} \GeV$ we find for the effective light neutrino mass matrix 
$m_\nu$ \\
\small
\begin{equation}
m_\nu\approx\left(
\begin{array}{ccc}
-2.4129 + i \; 7.2799 \cdot 10^{-2} & 
   2.4662 + i \; 1.0216 \cdot 10^{-4}&	
   2.7217 - i \; 1.7759 \cdot 10^{-2}\\
.	& -2.4181  - i \; 7.3181 \cdot 10^{-2}&
	4.5802 + i \; 4.5367 \cdot 10^{-2}\\
.	&. & -1.7962 - i \; 2.3591 \cdot 10^{-2}\\
\end{array}
\right) \cdot 10^{-2} \, \eV\; .
\label{matrixnu2}
\end{equation}
\normalsize
Apparently, some corrections to the lowest order matrix from the higher-dimensional operators 
are sizable, since the diagonal elements are no longer equal, but $|(m_\nu)_{22}| 
\approx 1.347 \, |(m_\nu)_{33}|$,  
and also the off-diagonal elements differ, {\it e.g.}
$|(m_\nu)_{12}| \approx  0.538 \, |(m_\nu)_{23}|$.  
This is a result of the interplay between
different relatively small corrections in the matrices $m_D$ and $M_{NS}$.  

The matrix $m_\nu$ in \Eqref{matrixnu2} 
yields the following mass squared differences and mixing 
parameters
\begin{eqnarray}\nonumber
&&\Delta m^2_\mathrm{21} = 7.58\, \cdot 10^{-5}\eV^2\;,\quad\Delta m^2_\mathrm{31} =  2.58 \, 
\cdot 10^{-3}\eV^2\;,\quad r = 0.0294 \, \;,\\ 
&&\theta_{12} = 31.4 ^\circ\;,\quad\theta_{13} = 1.14 ^\circ\;,\quad\theta_{23} =  46.5 ^\circ\; ,\\ \nonumber
&&\delta= 323.2 ^\circ\;,\quad\varphi_1= 299.3 ^\circ\;,\quad\varphi_2= 160.8^\circ\; ,
\end{eqnarray}
with the Dirac and Majorana phases defined according to~\cite{Antusch:2003kp}.
These results are within the  $2\sigma$ bounds of~\cite{Maltoni:2004ei}.  
Notice that the deviation of the angle $\theta_{12}$ from its TBM value is significant.
The masses of the light neutrinos are $m_1= 0.0437 \eV$, $m_2= 0.0446 \eV$ and $m_3 = 0.0670 \eV$.
Thus, the light neutrinos are normally ordered and have only a mild hierarchy. 
The sum of the masses, $\sum m_i= 0.155 \eV$, is below the cosmological bound~\cite{cosmobounds}.
In the numerical analysis, we did not consider RG corrections. 
Although they might lead to sizable corrections due to the large hierarchy in the
right-handed neutrino masses~\cite{Lindner:2005pk}, 
these corrections can be included in a redefinition of the coefficients 
of the higher-dimensional operators. 
Similarly, we neglect possible corrections to the mixing angles coming from the non-diagonality 
of the charged lepton mass matrix.
These corrections
to the $1-2$ mixing angle are of the order of the Cabibbo angle, 
whereas they are smaller for $\theta_{13}$ and $\theta_{23}$. 

In this numerical example we find for the masses of the right-handed neutrinos $\approx 
\{ 3.83 \TeV,$ $1.47 \cdot 10^{10} \GeV, 1.16 \cdot 10^{12} \GeV\}$. This mass spectrum 
differs from the estimate in the lowest order presented  
in \Secref{sec:Sigma81lowest}.
Especially the two lighter states are heavier than before. 
The difference can be partly attributed to the actual value of $1/|\tilde A + 2 
\tilde B|$ and 
$|\tilde A + 2 \tilde B|/|3 \tilde B (2 \tilde A + \tilde B)|$ and partly to the fact that we
included some next-to-leading order corrections to the matrix $M_{NS}$.

\subsection{Flavon Potential}
\label{sec:S81flavon}

The renormalizable part of the flavon potential is given by
\footnote{Possible terms containing GUT Higgs fields and the flavons $\chi$ and/or $\chi^{\star}$,
which are invariant under all symmetries of the model, are assumed to be fine-tuned in order to
keep the flavon and the GUT Higgs potential appropriately separated.}
\small
\begin{equation}
V_\chi(\chi_i)=M^2_{\chi} \sum_j |\chi_j|^2 + \left[\kappa e^{i \sigma} \sum_j \chi_j^3 +\hc 
\right]+ 
\lambda_1\sum_j |\chi_j|^4+\lambda_2\sum_{j< k}|\chi_j|^2|\chi_k|^2\; ,
\end{equation}
\normalsize
where $M_{\chi}$ is the mass parameter, $\lambda_{1,2}$ and $\kappa$ are real constants, 
and the phase $\sigma$ lies in the interval $[0, 2 \pi)$. In
order to analyze the flavon potential, we parameterize the fields $\chi_i$ 
as $\chi_i \equiv X_i e^{i \xi_i}$ with $X_i \geq 0$ and $\xi_i \in [0,2\pi)$. 
Then
\small
\begin{equation}
V_\chi(X_i,\,\xi_j)=M^2_{\chi} \sum_k X_k^2 + \lambda_1\sum_k X_k^4+\lambda_2\sum_{k< l}
X_k^2 X_l^2+ 2 \kappa\sum_k X_k^3 \cos\left(\sigma + 3\xi_k\right)\; .
\end{equation}
\normalsize
The extremization conditions for the VEVs
$\braket{X_1}$ and $\braket{\xi_1}$ read
\small
\begin{subequations}
\begin{align}
\frac{\partial V_\chi}{\partial X_1}&=2
X_1 \left( M^2_{\chi} +2 \lambda_1 X_1^2+ \lambda_2 X_2^2+ \lambda_2 X_3^2+3 \kappa
X_1 \cos\left(\sigma +3 \xi_1\right)\right)=0\label{eq:grad1}\\
\frac{\partial V_\chi}{\partial \xi_1}&=-6 \kappa
X_1^3 \sin\left(\sigma +3 \xi_1\right)=0\label{eq:grad4}\; .
\end{align}
\end{subequations}
\normalsize
The corresponding equations for $\braket{X_{2,3}}$ and $\braket{\xi_{2,3}}$  are obtained
by cyclic permutation in the index $i=1,2,3$.
\Eqref{eq:grad4} is solved by either a vanishing VEV $\braket{X_i}$ or by the
relation $3\braket{\xi_i}+\sigma = n_i\,\pi$ 
($n_i = 0, \pm 1, \pm 2, ...$). Note that $n_i$ can be chosen independently for 
the three $\xi_i$. 
This allows one to choose a vacuum configuration in which
two of the three fields $X_i$ (equivalent to $\chi_i$) have a vanishing VEV. Such a 
configuration is a good approximation to the vacuum structure used to accommodate the
charged fermion mass hierarchy, see \Eqref{eq:S81vev-hier}.
If we set $\braket{X_1}=\braket{X_2}=0$, $\braket{X_3}\neq0$ and 
require that the extremum is a minimum of the potential, 
we obtain the following  
\begin{equation}
\braket{X_3}=\frac{3 \kappa + \sqrt{9 \kappa^2 - 8 M^2_{\chi}
     \lambda_1}}{4\lambda_1}\;,\quad\braket{\xi_3}=-\frac{\sigma \pm\pi}{3}\label{eq:sol2}
\end{equation}
together with the condition $9 \kappa^2 \geq 8 M_\chi^2 \lambda_1$. 
The real  value of the VEVs of $\chi_i$  
assumed in our previous discussion can be achieved by a suitable choice of the 
phase $\sigma$.
The questions how to generate the VEVs $\langle \chi_{1,2} \rangle$ and how to
ensure the hierarchy of these VEVs are not answered in this paper, since this requires 
a careful study of the higher-dimensional operators with more than four flavons contributing
to the scalar potential.

\section{Summary}
\label{sec:conclusions}

In the context of $\SO(10)$ grand unified models we have presented a mechanism which
cancels partially or completely the hierarchy of the Dirac mass matrix in the light neutrino mass matrix $m_\nu$.  
The ingredients of the mechanism are:
$(i)$ the existence of three fermionic $\SO(10)$ singlets,
which realize the double seesaw mechanism and
$(ii)$ a discrete non-Abelian flavor symmetry
which restricts the form of the relevant matrices
$m_{D}$, $M_{NS}$ and $M_{SS}$,  so that the structures
of $m_{D}$ and $M_{NS}$ are correlated.
The charged fermion mass matrices are diagonal at leading order and the
hierarchy of the up-quarks can be reproduced.
The framework of such a model can be supersymmetric or not,
depending on the choice of the flavor group.

We discuss the two discrete groups $T_7$ and $\Sigma(81)$,
which lead to such a cancellation.
In the first case the hierarchy is partially cancelled, and the existence of the 
dominant $2-3$ block in the light neutrino mass matrix, responsible for 
maximal atmospheric mixing, is related to the geometric hierarchy of the 
up-quarks, $m_u:m_c:m_t = \epsilon^4:\epsilon^2:1$ for $\epsilon \approx 0.05$.
In the  $\Sigma(81)$ model we can  achieve a
complete cancellation of the up-quark hierarchy in $m_{\nu}$. 
The resulting $m_{\nu}$ is compatible with TBM, however the atmospheric
mass squared difference vanishes at leading order. In none of the two models
we can accommodate all features of the neutrino masses and lepton mixings
by considering only the leading order. Thus, further effects have to be taken into
account. It turns out that in the $T_7$ model the inclusion of a second Higgs
in the $\SO(10)$ representation $\overline{\Rep{16}}$ can produce a linear
seesaw contribution leading to a large $1-2$ mixing angle without disturbing
the cancellation mechanism from which $\theta_{23} \approx \frac{\pi}{4}$
originates. In the $\Sigma(81)$ model the situation is even simpler, since the
introduction of operators involving more than one flavon field already lifts
the degeneracy in the light neutrino mass spectrum and allows to reproduce 
the experimental data. 

Since  the top quark mass is also generated by non-renormalizable
operators, at least one of the expansion parameters
$\braket{\chi_{i}}/\Lambda$ has to be of order one. As a consequence,
contributions to the fermion masses from operators with more than one flavon 
are in general not small and have to be
carefully studied.
For the $T_{7}$ model we showed that an additional $Z_7$ symmetry
can forbid all operators which invalidate the predictions made at lowest order.   
Additionally, the introduction of the $Z_7$ symmetry helps to solve
the problem of a too large light neutrino mass scale by allowing
the singlets to acquire masses only through non-renormalizable operators
of the form $S S \chi^n/\Lambda^{n-1}$. In the $\Sigma(81)$ realization
we observe that the higher-dimensional operators have the same structure
 in the small expansion parameter $\epsilon \approx 0.05$
as the leading order terms.
These operators are therefore much less dangerous than in the $T_7$ model and
even help to produce the correct atmospheric mass squared difference.

The Cabibbo angle in the quark 
sector can be accommodated in both models, if additional operators of the 
structure as found in \Eqref{eq:thCoperatorT7} and \Eqref{16prime}
are introduced which only
contribute to the down quark and the charged lepton mass matrix. The key feature
is that the Higgs fields $\Rep{16}_H$, $\Rep{16}'_H$ have to transform
as non-trivial one-dimensional representations of the flavor group.  

Additionally, we study in both models the mass spectrum of the heavy neutral fermions. In case
of the $T_7$ model we find that these particles have masses between $10^{8} \GeV$ and $10^{15} \GeV$.
In contrast to that the right-handed neutrinos are strongly hierarchical in the
$\Sigma(81)$ case with the lightest right-handed neutrino being as light as $\sim 400 \GeV$ to
$\sim 4 \TeV$. The singlets, however, are much heavier with masses 
around the Planck scale. Although one right-handed neutrino is very light, its effects are practically unobservable due to the tiny coupling to the active neutrinos.

Finally, we compute the flavon (super-)potential of the models. In both cases the
vacuum structure can be well approximated  by a configuration in which only one of the 
three flavons has a non-vanishing VEV. Such a configuration can be a minimum in both
potentials. To study further whether the hierarchy among the other VEVs, which are zero
at leading order, can be appropriately produced we however would need to consider non-renormalizable
operators with several flavons also in the (super-)potential.

In conclusion, we consider $\SO(10)$ GUTs with three additional fermionic singlets
which can reconcile the different mass hierarchies in the charged fermion 
and the neutrino sector 
and explain the peculiar mixing pattern among the leptons with the help of discrete
non-Abelian flavor symmetries. Furthermore, we study several aspects of these models such
as the effects of higher-dimensional operators, the generation of the Cabibbo angle, the mass
spectrum of the heavy neutral fermions and the flavon potential at leading order.

\subsubsection*{Acknowledgements}
We would like to thank M.~Lindner who participated in the early stages
of this work. We also thank A.~Blum for discussions.
C.~H. and M.~S. acknowledge support from the Deutsche
Forschungsgemeinschaft in the Transregio Sonderforschungsbereich TR27
``Neutrinos and Beyond - Weakly interacting particles in Physics, Astrophysics and Cosmology''.
A.~Yu.~S. acknowledges support  by the Alexander von Humboldt Foundation.   
\newpage
\appendix

\mathversion{bold}
\section{Group Theory of \Groupname{T}{7}\label{app:T7}}
\mathversion{normal}

In this appendix we show the character table, Kronecker products, one set of
generators and the resulting Clebsch-Gordan coefficients for the group $T_7$.
As already mentioned above, this group is very similar to $A_4$ with the crucial
difference that the two three-dimensional representations of
\Groupname{T}{7} are complex and not real like the one of
\Groupname{A}{4}.\\
\begin{table}
\begin{center}
\begin{tabular}{l|ccccc|c}
&\multicolumn{5}{|c|}{classes}                                                   &\\ \cline{2-6}
&$\Cl{1}$&$\Cl{2}$&$\Cl{3}$&$\Cl{4}$&$\Cl{5}$&\\
\cline{1-6}
\rule[0.15in]{0cm}{0cm} $\rm G$  &$\rm \mathbb{1}$&$\rm B$&$\rm B ^{2}$&$\rm A$&$\rm A ^{3}$&\\
\cline{1-6}
$\OrdCl{i}$                  &1      &7              &7                       &3              &3&\\
\cline{1-6}
$\Ord{h}{\Cl{i}}$                           &1      &3              &3                       &7              &7      &$\mathbb{c} ^{(\mu)}$\\
\hline
$\MoreRep{1}{1}$                                &1      &1              &1                       &1              &1                              &1\\
$\MoreRep{1}{2}$                                &1      &$\omega$
&$\omega ^{2}$    &1              &1              &0\\
$\MoreRep{1}{3}$                                &1
&$\omega ^{2}$&$\omega$                   &1              &1              &0\\
$\Rep{3}$                                &3      &0              &0
&$\xi$          &$\xi ^{\star}$       &0\\
$\Rep{3} ^{\star}$                                &3      &0
&0                      &$\xi ^{\star}$       &$\xi$          &0\\[0.1cm]
\hline
\end{tabular}
\end{center}
\begin{center}
\begin{minipage}[t]{12cm}
\caption[]{Character table of $T_7$. $\omega \equiv e^{\frac{2 \pi i}{3}}$ and $\xi \equiv \frac{1}{2} (-1 + i
   \sqrt{7})$, so that $\xi= \rho + \rho^{2} + \rho^{4}$ where $\rho \equiv e^{\frac{2 \pi i}{7}}$.
   $\Cl{i}$ are the classes of the group, $\OrdCl{i}$ is the order of the $i ^{\mathrm{th}}$ class, 
{\it i.e.} the number of distinct elements contained in this class, $\Ord{h}{\Cl{i}}$
is the order of the elements $S$ in the class $\Cl{i}$, {\it i.e.} the smallest
integer ($>0$) for which the equation $S ^{\Ord{h}{\Cl{i}}}= \mathbb{1}$
holds. Furthermore the table contains one representative for each
class $\Cl{i}$ given as product of the generators $\rm
A$ and $\rm B$ of the group. Finally, $\mathbb{c} ^{(\mu)}$ denotes the so-called $\mathbb{c}$-value
of a representation $\mu$ which indicates whether $\mu$ is real ($\mathbb{c}^{(\mu)}=1$), pseudo-real
($\mathbb{c}^{(\mu)}=-1$) or complex ($\mathbb{c}^{(\mu)}=0$).
\label{tab:chtabT7}
}
\end{minipage}
\end{center}
\end{table}
The character table is presented in \Tabref{tab:chtabT7}. The used generators for the three-dimensional
representations are~\cite{generatorsT7}
\small
\begin{eqnarray}
\nonumber
&&\;\;\;\;\;\;\;\;\;\;\;\;\;\;\;\;\;
\Rep{3} \; \mbox{:} \; \rm A=\left(\begin{array}{ccc}
                                                 e^{\frac{2 \pi i}{7}} & 0 &  0 \\
                                                  0 & e^{\frac{4 \pi i}{7}} & 0 \\
                                                  0 & 0 & e^{\frac{8 \pi i }{7}}
                                 \end{array}\right) \; , \;\; \rm B=\left(\begin{array}{ccc}
                                                 0 & 1 & 0 \\
                                                  0 & 0 & 1 \\
                                                  1 & 0 & 0
                                 \end{array}\right)\\ \nonumber
&&\mbox{\normalsize
and
\small}\\
\nonumber
&&\;\;\;\;\;\;\;\;\;\;\;\;\;\;\;\;\;
\Rep{3} ^{\star} \; \mbox{:} \; \rm A=\left(\begin{array}{ccc}
                                                 e^{-\frac{2 \pi i }{7}} & 0  & 0 \\
                                                  0 & e^{-\frac{4 \pi i}{7}}  & 0 \\
                                                  0 & 0 & e^{-\frac{8 \pi i }{7}}
                                 \end{array}\right) \; , \;\; \rm B=\left(\begin{array}{ccc}
                                                 0 & 1 & 0 \\
                                                  0 & 0 & 1 \\
                                                  1 & 0 & 0
                                 \end{array}\right) \; .
\end{eqnarray}
\normalsize
They fulfill the relations~\cite{generatorsT7}
\small
\[
\rm A^7=\mathbb{1} \; , \;\; \rm B^3=\mathbb{1} \; , \;\; \rm AB=BA^4
\; .
\]
\normalsize
The Kronecker products are

\small
\parbox{2.5in}{
\begin{align*}
&\MoreRep{1}{1} \times \MoreRep{1}{i} = \MoreRep{1}{i} \; ,
\;\; \MoreRep{1}{2} \times
\MoreRep{1}{3} = \MoreRep{1}{1} \; , \\
&\MoreRep{1}{2} \times
\MoreRep{1}{2} = \MoreRep{1}{3} \; , \;\; \MoreRep{1}{3}
  \times \MoreRep{1}{3} = \MoreRep{1}{2} \; , \\
&\MoreRep{1}{i} \times \Rep{3}  = \Rep{3} \; , \;\;
\MoreRep{1}{i} \times \Rep{3} ^{\star} = \Rep{3} ^{\star} \; , \;\;
\end{align*}}
\parbox{2.5in}{
\begin{align*}
&\left[ \Rep{3} \times \Rep{3}\right] = \Rep{3} + \Rep{3} ^{\star} \; ,
\;\;\;\; \left\{ \Rep{3} \times \Rep{3}\right\} = \Rep{3} ^{\star} \; ,\\
&\left[ \Rep{3} ^{\star} \times \Rep{3} ^{\star} \right] = \Rep{3} +
\Rep{3} ^{\star} \; , \;\;\;\; \left\{ \Rep{3}^{\star} \times \Rep{3}^{\star} \right\} = \Rep{3} \; ,\\
&\Rep{3} \times \Rep{3} ^{\star} = \MoreRep{1}{1} + \MoreRep{1}{2} +
\MoreRep{1}{3} + \Rep{3} + \Rep{3} ^{\star}
\end{align*}}\\
\normalsize
with $\left[ \mu \times \mu \right]$ being the symmetric part of the
product $\mu \times \mu$ and $\left\{ \mu \times \mu \right\}$ being
the anti-symmetric part.
The non-trivial Clebsch-Gordan coefficients
for $\left(a_{1}, a_{2}, a_{3} \right) ^{T} \sim \Rep{3}$, $\left(b_{1}, b_{2}, b_{3} \right) ^{T} \sim \Rep{3} ^{\star}$ and $c
\sim \MoreRep{1}{1}$, $c ^{\prime} \sim \MoreRep{1}{2}$, $c ^{\prime\prime} \sim \MoreRep{1}{3}$ are\\
\small
\hspace{-0.6in}\parbox{3in}{
\begin{align*}
&\Rep{3} \times \MoreRep{1}{1} \;\; \mbox{:} \;\; \left( a_{1} \, c, a_{2} \,
   c , a_{3} \, c \right)^{T} \sim \Rep{3} \\
&\Rep{3} \times \MoreRep{1}{2} \;\; \mbox{:} \;\; \left(a_{1} \, c ^{\prime},
   \omega \, a_{2} \, c^{\prime} , \omega ^{2} \, a_{3} \, c ^{\prime} \right)^{T} \sim \Rep{3} \\
&\Rep{3} \times \MoreRep{1}{3} \;\; \mbox{:} \;\;
\left(a_{1} \, c^{\prime \, \prime} , \omega ^{2} \, a_{2} \,
   c^{\prime \, \prime} , \omega \, a_{3} \, c ^{\prime \, \prime} \right)^{T} \sim
\Rep{3} \\
\end{align*}}
\parbox{3in}{
\begin{align*}
&\Rep{3} ^{\star} \times \MoreRep{1}{1} \;\; \mbox{:} \;\; \left( b_{1}\, c,
   b_{2} \, c , b_{3} \, c \right)^{T} \sim
\Rep{3} ^{\star} \\
&\Rep{3} ^{\star} \times \MoreRep{1}{2} \;\; \mbox{:} \;\;
\left(b_{1} \, c^{\prime} , \omega \, b_{2} \, c^{\prime} , \omega ^{2} \, b_{3} \, c^{\prime}
\right)^{T} \sim \Rep{3} ^{\star} \\
&\Rep{3} ^{\star} \times \MoreRep{1}{3} \;\; \mbox{:} \;\;
\left(b_{1} \, c^{\prime \, \prime} , \omega^2 \, b_{2} \, c^{\prime
     \, \prime} , \omega \, b_{3} \, c^{\prime \, \prime}   \right)^{T} \sim
\Rep{3} ^{\star} \\
\end{align*}
\normalsize
}

\noindent For $\left( a_{1}, a_{2}, a_{3}\right) ^{T},\left( a_{1} ^{\prime},
   a_{2} ^{\prime}, a_{3} ^{\prime} \right) ^{T} \sim \Rep{3}$ we find
\small
\begin{equation}
\nonumber
  \left( a_{3} \, a_{3} ^{\prime}, a_{1} \, a_{1} ^{\prime}, a_{2} \,
    a_{2} ^{\prime} \right) ^{T}  \sim \Rep{3} \; , \;\;
  \left( a_{2} \, a_{3} ^{\prime}, a_{3} \, a_{1} ^{\prime}, a_{1} \,
    a_{2} ^{\prime} \right) ^{T} \sim \Rep{3} ^{\star} \;\; \mbox{and} \;\;
   \left( a_{3} \, a_{2} ^{\prime}, a_{1} \, a_{3} ^{\prime}, a_{2} \,
     a_{1} ^{\prime} \right) ^{T} \sim \Rep{3}^{\star}
\end{equation}
\normalsize

\noindent and for $\left( b_{1}, b_{2}, b_{3} \right) ^{T},\left( b_{1} ^{\prime},
   b_{2} ^{\prime}, b_{3} ^{\prime} \right) ^{T} \sim \Rep{3} ^{\star}$
\small
\begin{equation*}
  \left( b_{2} \, b_{3} ^{\prime}, b_{3} \, b_{1} ^{\prime}, b_{1} \,
    b_{2} ^{\prime} \right) ^{T}  \sim \Rep{3}\; , \;\;
  \left( b_{3} \, b_{2} ^{\prime}, b_{1} \, b_{3} ^{\prime}, b_{2} \,
    b_{1} ^{\prime} \right) ^{T} \sim \Rep{3} \;\; \mbox{and} \;\;
   \left( b_{3} \, b_{3} ^{\prime}, b_{1} \, b_{1} ^{\prime}, b_{2} \,
     b_{2} ^{\prime} \right) ^{T} \sim \Rep{3}^{\star} \; .
\end{equation*}
\normalsize

\noindent For $\left(a_{1}, a_{2}, a_{3} \right) ^{T} \sim \Rep{3}$,
$\left(b_{1}, b_{2}, b_{3} \right) ^{T}  \sim \Rep{3} ^{\star}$ the
\Groupname{T}{7} covariant combinations are
\small
\begin{align*}
&a_{1} \, b_{1} + a_{2} \, b_{2} + a_{3} \, b_{3} \sim \MoreRep{1}{1}\; , \;\;
  a_{1} \, b_{1} + \omega^{2} \, a_{2} \, b_{2} + \omega \, a_{3} \,
  b_{3} \sim \MoreRep{1}{2}\; , \;\; 
a_{1} \, b_{1} + \omega \, a_{2} \, b_{2} + \omega^{2} \, a_{3} \, b_{3} \sim
\MoreRep{1}{3}\; ,\\
& \;\;\;\;\;\;\;\;\;\;\;\;\;\;\;\;\;\;\;\;\;\;\;\;\;\;
\left( a_{2} \, b_{1}, a_{3} \, b_{2}, a_{1} \, b_{3} \right) ^{T}
\sim \Rep{3} \;\;\; \mbox{and} \;\;\; \left( a_{1} \, b_{2}, a_{2} \, b_{3}, a_{3} \,
   b_{1} \right) ^{T} \sim \Rep{3}^{\star} \; .
\end{align*}
\normalsize

\mathversion{bold}
\section{Group Theory of $\Sigma (81)$\label{app:Sigma81}}
\mathversion{normal}

The irreducible representations are $\MoreRep{1}{i}$ with $\rm i=1,...,9$
and $\MoreRep{3}{i}$ with $\rm i=1,...,8$. Apart from $\MoreRep{1}{1}$ which is the total singlet
all other representations are complex. The complex conjugated pairs are found in
\Tabref{tab:conjrep_Sigma81}. Six of the eight three-dimensional representations are faithful.

\begin{table}
\begin{center}
\begin{tabular}{|c|c|cccc|cccc|}
\hline
rep. & $\MoreRep{1}{1}$ & $\MoreRep{1}{2}$  & $\MoreRep{1}{4}$
&$\MoreRep{1}{5}$ & $\MoreRep{1}{6}$
&$\MoreRep{3}{1}$ &$\MoreRep{3}{3}$ & $\MoreRep{3}{5}$ &$\MoreRep{3}{7}$ \\
\hline
$\mathrm{rep.} ^{\star}$ & $\MoreRep{1}{1}$ & $\MoreRep{1}{3}$
&$\MoreRep{1}{7}$ &$\MoreRep{1}{8}$ &$\MoreRep{1}{9}$
& $\MoreRep{3}{2}$& $\MoreRep{3}{4}$ &$\MoreRep{3}{6}$  &$\MoreRep{3}{8}$ \\
\hline
\end{tabular}
\end{center}
\begin{center}
\begin{minipage}[t]{12cm}
\caption[]{The representations of the group $\Sigma(81)$ and their complex conjugates.
\label{tab:conjrep_Sigma81}}
\end{minipage}
\end{center}
\end{table}

The character table is given in~\cite{Ma:2006ht}
together with a choice of representation matrices for the
representation $\MoreRep{3}{1}$ which is called \mathversion{bold} $\rm 3_A$
\mathversion{normal} in~\cite{Ma:2006ht}. For the other representations of 
$\Sigma(81)$ we have the following identifications: we use the same notation
for all one-dimensional representations $\MoreRep{1}{i}$ and for the triplets
we have \mathversion{bold} $\rm \bar{3}_A$ \mathversion{normal} $\equiv \MoreRep{3}{2}$,
 \mathversion{bold} $\rm 3_B$ \mathversion{normal} $\equiv \MoreRep{3}{3}$,
 \mathversion{bold} $\rm \bar{3}_B$ \mathversion{normal} $\equiv \MoreRep{3}{4}$,
 \mathversion{bold} $\rm 3_C$ \mathversion{normal} $\equiv \MoreRep{3}{5}$,
 \mathversion{bold} $\rm\bar{3}_C$ \mathversion{normal} $\equiv \MoreRep{3}{6}$,
 \mathversion{bold} $\rm 3_D$ \mathversion{normal} $\equiv \MoreRep{3}{7}$,
 and \mathversion{bold} $\rm\bar{3}_D$ \mathversion{normal} $\equiv \MoreRep{3}{8}$.

The generators of all representations are presented in \Tabref{tab:genSigma81}.

Some of the Kronecker products are already shown in~\cite{Ma:2006ht,Ma:2007ku}. In
\Tabref{tab:KroneckerSigma81} we show the products which we
need to discuss the lowest order.

\begin{table}
\begin{center}
\subtable[Kronecker products with one-dimensional representations]{\begin{tabular}{c|ccc}
rep. & $\MoreRep{1}{1}$ & $\MoreRep{1}{2}$ & $\MoreRep{1}{3}$ \\\hline
$\MoreRep{1}{1}$ & $\MoreRep{1}{1}$ & $\MoreRep{1}{2}$ & $\MoreRep{1}{3}$\\
$\MoreRep{1}{2}$ &  $\MoreRep{1}{2}$&  $\MoreRep{1}{3}$&  $\MoreRep{1}{1}$\\
$\MoreRep{1}{3}$ &  $\MoreRep{1}{3}$&  $\MoreRep{1}{1}$&  $\MoreRep{1}{2}$\\\hline
$\MoreRep{3}{1}$ & $\MoreRep{3}{1}$ & $\MoreRep{3}{1}$ & $\MoreRep{3}{1}$ \\
$\MoreRep{3}{2}$ &  $\MoreRep{3}{2}$&  $\MoreRep{3}{2}$&  $\MoreRep{3}{2}$\\
\end{tabular}}
\hspace{1cm}
\subtable[Kronecker products of three-dimensional representations]{\begin{minipage}{3in}
\begin{eqnarray}\nonumber
\left[ \MoreRep{3}{1} \times \MoreRep{3}{1} \right] = \MoreRep{3}{2} + \MoreRep{3}{4}
\;\;\; &\mbox{and}& \;\;\;
\left\{ \MoreRep{3}{1} \times \MoreRep{3}{1} \right\} = \MoreRep{3}{4} \; ,
\\ \nonumber
\left[ \MoreRep{3}{2} \times \MoreRep{3}{2} \right] = \MoreRep{3}{1} + \MoreRep{3}{3}
\;\;\; &\mbox{and}& \;\;\;
\left\{ \MoreRep{3}{2} \times \MoreRep{3}{2} \right\} = \MoreRep{3}{3} \; ,
\end{eqnarray}
\begin{eqnarray}\nonumber
\MoreRep{3}{1} \times \MoreRep{3}{2} &=& \MoreRep{1}{1} + \MoreRep{1}{2} + \MoreRep{1}{3}
+ \MoreRep{3}{7} + \MoreRep{3}{8} \; .
\end{eqnarray}
\end{minipage}
}
\begin{center}
\begin{minipage}[t]{16cm}
\caption[]{Kronecker products of $\Sigma(81)$ relevant for the discussion of the leading order of the fermion
 masses. Note that $\left[\mu \times \mu \right]$ denotes the
symmetric and $\left\{ \mu \times \mu \right\}$ the anti-symmetric part of the product $\mu \times \mu$.
\label{tab:KroneckerSigma81}}
\end{minipage}
\end{center}
\end{center}
\end{table}

The non-trivial Clebsch-Gordan coefficients\footnote{The remaining
   ones can be obtained via the formulae given
in~\cite{calcCGC}.
}
are for $\left(a_{1}, a_{2}, a_{3} \right) ^{T} \sim \MoreRep{3}{i}$ ($\rm i=1,2$), $c
\sim \MoreRep{1}{1}$, $c^{\prime} \sim \MoreRep{1}{2}$ and  $c^{\prime\prime} \sim \MoreRep{1}{3}$
\begin{eqnarray}
\nonumber
&& \MoreRep{3}{i} \times \MoreRep{1}{1} \;\; \mbox{:} \;\; \left( a_{1} \, c, a_{2} \,
   c , a_{3} \, c \right)^{T} \sim \MoreRep{3}{i} \\ \nonumber
&& \MoreRep{3}{i} \times \MoreRep{1}{2} \;\; \mbox{:} \;\; \left(a_{1} \, c^\prime,
   \omega \, a_{2} \, c^\prime, \omega ^{2} \, a_{3} \, c^\prime \right)^{T} \sim \MoreRep{3}{i} \\ \nonumber
&& \MoreRep{3}{i} \times \MoreRep{1}{3} \;\; \mbox{:} \;\; \left(a_{1} \, c^{\prime\prime},
   \omega ^2 \, a_{2} \, c^{\prime\prime}, \omega \, a_{3} \, c^{\prime\prime} \right)^{T} \sim \MoreRep{3}{i} \; .
\end{eqnarray}

\noindent For $\left( a_{1}, a_{2}, a_{3}\right) ^{T}, \left( a_{1}^{\prime},
   a_{2}^{\prime}, a_{3}^{\prime} \right) ^{T} \sim \MoreRep{3}{1}$ the structure of the 
Clebsch-Gordan coefficients is
\begin{subequations}
\small
\begin{gather} \nonumber
  \left( a_{1} \, a_{1}^{\prime}, a_{2} \, a_{2}^{\prime}, a_{3} \, a_{3}^{\prime} \right) ^{T}
  \sim \MoreRep{3}{2}\; ,\quad
  \left( a_{2} \, a_{3}^{\prime}, a_{3} \, a_{1}^{\prime}, a_{1} \, a_{2}^{\prime} \right) ^{T}  
\sim \MoreRep{3}{4}\; ,\quad
  \left( a_{3} \, a_{2}^{\prime}, a_{1} \, a_{3}^{\prime}, a_{2} \, a_{1}^{\prime} \right) ^{T}  
\sim \MoreRep{3}{4} \; .
\end{gather}
\normalsize
\end{subequations}

\noindent Similarly, for $\left( b_{1}, b_{2}, b_{3}\right) ^{T}, \left( b_{1}^{\prime},
   b_{2}^{\prime}, b_{3}^{\prime} \right) ^{T} \sim \MoreRep{3}{2}$ we find
\begin{subequations}
\small
\begin{gather} \nonumber
  \left( b_{1} \, b_{1}^{\prime}, b_{2} \, b_{2}^{\prime}, b_{3} \, b_{3}^{\prime} \right) ^{T}  \sim \MoreRep{3}{1}\; ,\quad
  \left( b_{2} \, b_{3}^{\prime}, b_{3} \, b_{1}^{\prime}, b_{1} \, b_{2}^{\prime} \right) ^{T}  \sim \MoreRep{3}{3}\; ,\quad
  \left( b_{3} \, b_{2}^{\prime}, b_{1} \, b_{3}^{\prime}, b_{2} \, b_{1}^{\prime} \right) ^{T}  \sim \MoreRep{3}{3} \; .
\end{gather}
\normalsize
\end{subequations}

\noindent For  $\left( a_{1}, a_{2}, a_{3}\right) ^{T} \sim \MoreRep{3}{1}$  and $\left( b_{1},
   b_{2}, b_{3} \right) ^{T} \sim \MoreRep{3}{2}$ the covariant combinations read
\begin{subequations}
\small
\begin{gather} \nonumber
a_{1} \, b_{1} + a_{2} \, b_{2} + a_{3} \, b_{3}  \sim \MoreRep{1}{1}\; ,\quad
a_{1} \, b_{1} + \omega^2 \, a_{2} \, b_{2} + \omega \, a_{3} \, b_{3}
\sim \MoreRep{1}{2}\; ,
a_{1} \, b_{1} + \omega \, a_{2} \, b_{2} + \omega^2 \, a_{3} \, b_{3}
\sim \MoreRep{1}{3}\; ,\\ \nonumber
\left( a_{3} \, b_{2},a_{2} \, b_{1},a_{1} \, b_{3} \right)^T \sim \MoreRep{3}{7} \;\;\; \mbox{and} \;\;\;
\left( a_{2} \, b_{3}, a_{1} \, b_{2},a_{3} \, b_{1} \right)^T
\sim \MoreRep{3}{8}\; .
\end{gather}
\normalsize
\end{subequations}
\begin{table}[ht!]
\begin{center}
\begin{tabular}{|c|ccc|}
\hline
rep. & $\rm A$ & $\rm B$ & $\rm C$\\
\hline
$\MoreRep{1}{1}$ & $1$ & $1$ & $1$\\
$\MoreRep{1}{2}$ & $\omega$ & $1$ & $1$\\
$\MoreRep{1}{3}$ & $\omega^2$ & $1$ & $1$\\
$\MoreRep{1}{4}$ & $1$ & $1$ & $\omega^2$\\
$\MoreRep{1}{5}$ & $\omega^2$ & $1$ & $\omega^2$\\
$\MoreRep{1}{6}$ & $\omega$ & $1$ & $\omega^2$\\
$\MoreRep{1}{7}$ & $1$ & $1$ & $\omega$\\
$\MoreRep{1}{8}$ & $\omega$ & $1$ & $\omega$\\
$\MoreRep{1}{9}$ & $\omega^2$ & $1$ & $\omega$\\
$\MoreRep{3}{1}$ & $\left(\begin{array}{ccc}
	0 & 1 & 0\\
	0 & 0 & 1\\
	1 & 0 & 0\\
	   \end{array}
	   \right)$&
	    $\left(\begin{array}{ccc}
	1 & 0 & 0\\
	0 & \omega & 0\\
	0 & 0 & \omega^2\\
	   \end{array}
	   \right)$&
	  $\left(\begin{array}{ccc}
	1 & 0 & 0\\
	0 & 1 & 0\\
	0 & 0 & \omega\\
	   \end{array}
	   \right)$\\
$\MoreRep{3}{2}$ & $\left(\begin{array}{ccc}
	0 & 1 & 0\\
	0 & 0 & 1\\
	1 & 0 & 0\\
	   \end{array}
	   \right)$&
	    $\left(\begin{array}{ccc}
	1 & 0 & 0\\
	0 & \omega^2 & 0\\
	0 & 0 & \omega\\
	   \end{array}
	   \right)$&
	  $\left(\begin{array}{ccc}
	1 & 0 & 0\\
	0 & 1 & 0\\
	0 & 0 & \omega^2\\
	   \end{array}
	   \right)$\\
$\MoreRep{3}{3}$ & $\left(\begin{array}{ccc}
	0 & 1 & 0\\
	0 & 0 & 1\\
	1 & 0 & 0\\
	   \end{array}
	   \right)$&
	    $\left(\begin{array}{ccc}
	1 & 0 & 0\\
	0 & \omega & 0\\
	0 & 0 & \omega^2\\
	   \end{array}
	   \right)$&
	  $\left(\begin{array}{ccc}
	\omega^2 & 0 & 0\\
	0 & \omega^2 & 0\\
	0 & 0 & 1\\
	   \end{array}
	   \right)$\\
$\MoreRep{3}{4}$ & $\left(\begin{array}{ccc}
	0 & 1 & 0\\
	0 & 0 & 1\\
	1 & 0 & 0\\
	   \end{array}
	   \right)$&
	    $\left(\begin{array}{ccc}
	1 & 0 & 0\\
	0 & \omega^2 & 0\\
	0 & 0 & \omega\\
	   \end{array}
	   \right)$&
	  $\left(\begin{array}{ccc}
	\omega & 0 & 0\\
	0 & \omega & 0\\
	0 & 0 & 1\\
	   \end{array}
	   \right)$\\
$\MoreRep{3}{5}$ & $\left(\begin{array}{ccc}
	0 & 1 & 0\\
	0 & 0 & 1\\
	1 & 0 & 0\\
	   \end{array}
	   \right)$&
	    $\left(\begin{array}{ccc}
	1 & 0 & 0\\
	0 & \omega & 0\\
	0 & 0 & \omega^2\\
	   \end{array}
	   \right)$&
	  $\left(\begin{array}{ccc}
	\omega & 0 & 0\\
	0 & \omega & 0\\
	0 & 0 & \omega^2\\
	   \end{array}
	   \right)$\\
$\MoreRep{3}{6}$ & $\left(\begin{array}{ccc}
	0 & 1 & 0\\
	0 & 0 & 1\\
	1 & 0 & 0\\
	   \end{array}
	   \right)$&
	    $\left(\begin{array}{ccc}
	1 & 0 & 0\\
	0 & \omega^2 & 0\\
	0 & 0 & \omega\\
	   \end{array}
	   \right)$&
	  $\left(\begin{array}{ccc}
	\omega^2 & 0 & 0\\
	0 & \omega^2 & 0\\
	0 & 0 & \omega\\
	   \end{array}
	   \right)$\\
$\MoreRep{3}{7}$ & $\left(\begin{array}{ccc}
	0 & 0 & 1\\
	1 & 0 & 0\\
	0 & 1 & 0\\
	   \end{array}
	   \right)$&
	    $\left(\begin{array}{ccc}
	\omega & 0 & 0\\
	0 & \omega & 0\\
	0 & 0 & \omega\\
	   \end{array}
	   \right)$&
	  $\left(\begin{array}{ccc}
	\omega & 0 & 0\\
	0 & 1 & 0\\
	0 & 0 & \omega^2\\
	   \end{array}
	   \right)$\\
$\MoreRep{3}{8}$ & $\left(\begin{array}{ccc}
	0 & 0 & 1\\
	1 & 0 & 0\\
	0 & 1 & 0\\
	   \end{array}
	   \right)$&
	    $\left(\begin{array}{ccc}
	\omega^2 & 0 & 0\\
	0 & \omega^2 & 0\\
	0 & 0 & \omega^2\\
	   \end{array}
	   \right)$&
	  $\left(\begin{array}{ccc}
	\omega^2 & 0 & 0\\
	0 & 1 & 0\\
	0 & 0 & \omega\\
	   \end{array}
	   \right)$\\
\hline
\end{tabular}
\end{center}
\begin{center}
\begin{minipage}[t]{13.5cm}
\caption[]{Generators of $\Sigma (81)$. We show three generators $\rm A$, $\rm B$ and $\rm C$ for
each representation, although it is enough to take the generators
$\rm A$ and $\rm C$ in order to reproduce the whole group. Note that
$\omega \equiv e ^{\frac{2 \, \pi \, i}{3}}$.\label{tab:genSigma81}}
\end{minipage}
\end{center}
\end{table}

\newpage
\section{Higher-Dimensional Operators}
\label{app:higherdim_operators}
\newpage
\thispagestyle{empty}
\begin{landscape}
\begin{table}
\scriptsize
\hspace{-1.3cm}\begin{tabular}{|c|c|c|c|c|c|c|}
\hline
Order & \multicolumn{6}{|c|}{Operators}\\
\hline
& $\Order{1}$ &  $\Order{\epsilon^{4}}$ &  $\Order{\epsilon^{2}}$
& $\Order{\epsilon^{6}}$ & $\Order{\epsilon^{4}}$ &  $\Order{\epsilon^{6}}$\\
\cline{2-7}
& $\chi_{1} ^{n}$ & $\chi_{1} ^{n-1} \, \chi_{2}$ & $\chi_{1} ^{n-1} \, \chi_{3}$ &
$\chi_{1} ^{n-2} \, \chi_{2} \, \chi_{3}$ & $\chi_{1} ^{n-2} \, \chi_{3} ^{2}$
& $\chi_{1} ^{n-3} \, \chi_{3} ^{3}$\\
\hline \hline
$n=1$ & \multicolumn{3}{|c|}{$\left( \begin{array}{c}
	\chi_{1}\\
	\chi_{2}\\
	\chi_{3}
	\end{array} \right) \sim \Rep{3} ^{\star}$} & & & \\
\hline
$n=2$	& $\left( \begin{array}{c}
	\chi_{3} ^{2}\\
	\chi_{1} ^{2}\\
	\\
	\end{array} \right) \sim \Rep{3} ^{\star}$ &
	\multicolumn{3}{|c|}{$\left( \begin{array}{c}
	\chi_{2} \, \chi_{3}\\
	\chi_{1} \, \chi_{3}\\
	\chi_{1} \, \chi_{2}
	\end{array} \right) \sim \Rep{3}$}  & see $\chi_{1}^{n}$ & \\
\hline
$n=3$	& $\left( \begin{array}{c}
	\\
	\chi_{3} ^{3}\\
	\chi_{1} ^{3}
	\end{array} \right) \sim \Rep{3}$ &
	$\left( \begin{array}{c}
	\\
	\chi_{1} \, \chi_{3} ^{2}\\
	\chi_{1} ^{2} \, \chi_{2}
	\end{array} \right) \sim \Rep{3} ^{\star}$ &
	$\left( \begin{array}{c}
	\chi_{1} ^{2} \, \chi_{3}\\
	\\
	\\
	\end{array} \right) \sim \Rep{3}$ &
	$\chi_{1} \, \chi_{2} \, \chi_{3} \sim \MoreRep{1}{1}$
	& see $\chi_{1}^{n-1} \, \chi_{2}$
	& see $\chi_{1} ^{n}$\\
\hline
$n=4$	& $\left( \begin{array}{c}
	\\
	\\
	\chi_{1} ^{4}
	\end{array} \right) \sim \Rep{3} ^{\star}$ &
	$\left( \begin{array}{c}
	\chi_{1} \, \chi_{3} ^{3}\\
	\chi_{1} ^{3} \, \chi_{2}\\
	\\
	\end{array} \right) \sim \Rep{3}$ &
	$\chi_{1}^{3} \, \chi_{3} + \chi_{1} \, \chi_{2}^{3}
	+ \chi_{2} \, \chi_{3} ^{3} \sim \MoreRep{1}{1}$ &
	$\left( \begin{array}{c}
	\chi_{1} ^{2} \, \chi_{2} \, \chi_{3}\\
	\\
	\\
	\end{array} \right) \sim \Rep{3} ^{\star}$ &
	$\left( \begin{array}{c}
	\\
	\\
	\chi_{1}^{2} \, \chi_{3} ^{2}
	\end{array} \right) \sim \Rep{3}$
	& see $\chi_{1} ^{n-1} \, \chi_{2}$\\
\hline
$n=5$ 	&$\left( \begin{array}{c}
	\\
	\chi_{1} ^{5}\\
	\\
	\end{array} \right) \sim \Rep{3}$ &
	$\left( \begin{array}{c}
	\chi_{1} ^{4} \, \chi_{2}\\
	\\
	\\
	\end{array} \right) \sim \Rep{3}$ &
	$\left( \begin{array}{c}
	\chi_{1}^{4} \, \chi_{3}\\
	\\
	\\
	\end{array} \right) \sim \Rep{3} ^{\star}$ &
	$\left( \begin{array}{c}
	\\
	\chi_{1} ^{3} \, \chi_{2} \, \chi_{3}\\
	\\
	\end{array} \right) \sim \Rep{3} ^{\star}$ &
	$\left( \begin{array}{c}
	\\
	\\
	\chi_{1} ^{3} \, \chi_{3} ^{2}\\
	\end{array} \right) \sim \Rep{3} ^{\star}$ &
	$\chi_{1} ^{2} \, \chi_{3} ^{3} + \chi_{1}^{3} \, \chi_{2} ^{2}
	+ \chi_{2} ^{3} \, \chi_{3} ^{2} \sim \MoreRep{1}{1}$\\
\hline
$n=6$ 	&$\left( \begin{array}{c}
	\chi_{1} ^{6}\\
	\\
	\\
	\end{array} \right) \sim \Rep{3}$ &
	$\chi_{1} ^{5} \, \chi_{2} + \chi_{2} ^{5} \, \chi_{3}
	+ \chi_{1} \, \chi_{3} ^{5} \sim \MoreRep{1}{1}$ &
	$\left( \begin{array}{c}
	\\
	\chi_{1} ^{5} \, \chi_{3}\\
	\\
	\end{array} \right) \sim \Rep{3} ^{\star}$ &
	$\left( \begin{array}{c}
	\\
	\\
	\chi_{1} ^{4} \, \chi_{2} \, \chi_{3}\\
	\end{array} \right) \sim \Rep{3}$ &
	$\left( \begin{array}{c}
	\\
	\chi_{1} ^{4} \, \chi_{3} ^{2}\\
	\\
	\end{array} \right) \sim \Rep{3}$ &
	$\left( \begin{array}{c}
	\chi_{1} ^{3} \, \chi_{3} ^{3}\\
	\\
	\\
	\end{array} \right) \sim \Rep{3} ^{\star}$\\
\hline
$n=7$ 	& $\chi_{1} ^{7} + \chi_{2} ^{7} + \chi_{3} ^{7} \sim \MoreRep{1}{1}$ &
	$\left( \begin{array}{c}
	\chi_{1} ^{6} \, \chi_{2}\\
	\\
	\\
	\end{array} \right) \sim \Rep{3} ^{\star}$ &
	$\left( \begin{array}{c}
	\\
	\\
	\chi_{1} ^{6} \, \chi_{3}
	\end{array} \right) \sim \Rep{3}$ &
	$\left( \begin{array}{c}
	\\
	\\
	\chi_{1} ^{5} \, \chi_{2} \, \chi_{3}
	\end{array} \right) \sim \Rep{3} ^{\star}$ &
	$\left( \begin{array}{c}
	\chi_{1} ^{5} \, \chi_{3} ^{2}\\
	\\
	\\
	\end{array} \right) \sim \Rep{3}$ &
	$\left( \begin{array}{c}
	\\
	\chi_{1} ^{4} \, \chi_{3} ^{3}\\
	\\
	\end{array} \right) \sim \Rep{3} ^{\star}$\\
\hline
$n=8$	& $\left( \begin{array}{c}
	\chi_{1} ^{8}\\
	\\
	\\
	\end{array} \right) \sim \Rep{3} ^{\star}$ &
	$\left( \begin{array}{c}
	\\
	\chi_{1} ^{7} \, \chi_{2}\\
	\\
	\end{array} \right) \sim \Rep{3} ^{\star}$ &
	$\left( \begin{array}{c}
	\\
	\\
	\chi_{1} ^{7} \, \chi_{3}\\
	\end{array} \right) \sim \Rep{3} ^{\star}$ &
	$\left( \begin{array}{c}
	\\
	\chi_{1} ^{6} \, \chi_{2} \, \chi_{3}\\
	\\
	\end{array} \right) \sim \Rep{3}$ &
	$\chi_{1} ^{6} \, \chi_{3} ^{2} + \chi_{1} ^{2} \, \chi_{2} ^{6}\
	+ \chi_{2} ^{2} \, \chi_{3} ^{6} \sim \MoreRep{1}{1}$ &
	$\left( \begin{array}{c}
	\\
	\\
	\chi_{1} ^{5} \, \chi_{3} ^{3}
	\end{array} \right) \sim \Rep{3}$\\
\hline
$n=9$	& $\left( \begin{array}{c}
	\\
	\chi_{1} ^{9}\\
	\\
	\end{array} \right) \sim \Rep{3} ^{\star}$ &
	$\left( \begin{array}{c}
	\\
	\\
	\chi_{1} ^{8} \, \chi_{2}
	\end{array} \right) \sim \Rep{3}$ &
	$\left( \begin{array}{c}
	\\
	\chi_{1} ^{8} \, \chi_{3}\\
	\\
	\end{array} \right) \sim \Rep{3}$ &
	$\left( \begin{array}{c}
	\chi_{1} ^{7} \, \chi_{2} \, \chi_{3}\\
	\\
	\\
	\end{array} \right) \sim \Rep{3}$ &
	$\left( \begin{array}{c}
	\chi_{1} ^{7} \, \chi_{3} ^{2}\\
	\\
	\\
	\end{array} \right) \sim \Rep{3} ^{\star}$ &
	$\left( \begin{array}{c}
	\\
	\\
	\chi_{1} ^{6} \, \chi_{3} ^{3}
	\end{array} \right) \sim \Rep{3} ^{\star}$\\
\hline
$n=10$	&  $\left( \begin{array}{c}
	\\
	\\
	\chi_{1}^{10}
	\end{array} \right) \sim \Rep{3}$ &
	 $\left( \begin{array}{c}
	\\
	\\
	\chi_{1} ^{9} \, \chi_{2}
	\end{array} \right) \sim \Rep{3} ^{\star}$ &
	 $\left( \begin{array}{c}
	\chi_{1}^{9} \, \chi_{3}\\
	\\
	\\
	\end{array} \right) \sim \Rep{3}$ &
	$\chi_{1}^{8} \, \chi_{2} \, \chi_{3} + \chi_{1} \, \chi_{2} ^{8} \, \chi_{3}
	+ \chi_{1} \, \chi_{2} \, \chi_{3}^{8} \sim \MoreRep{1}{1}$ &
	 $\left( \begin{array}{c}
	\\
	\chi_{1} ^{8} \, \chi_{3}^{2}\\
	\\
	\end{array} \right) \sim \Rep{3} ^{\star}$ &
	 $\left( \begin{array}{c}
	\\
	\chi_{1} ^{7} \, \chi_{3}^{3}\\
	\\
	\end{array} \right) \sim \Rep{3}$ \\
\hline
$n=11$ 	& $\left( \begin{array}{c}
	\\
	\\
	\chi_{1} ^{11}\\
	\end{array} \right) \sim \Rep{3} ^{\star}$ &
	 $\left( \begin{array}{c}
	\\
	\chi_{1} ^{10} \, \chi_{2}\\
	\\
	\end{array} \right) \sim \Rep{3}$ &
	$\chi_{1} ^{10} \, \chi_{3} + \chi_{1} \, \chi_{2} ^{10}
	+ \chi_{2} \, \chi_{3} ^{10} \sim \MoreRep{1}{1}$ &
	 $\left( \begin{array}{c}
	\chi_{1}^{9} \, \chi_{2} \, \chi_{3}\\
	\\
	\\
	\end{array} \right) \sim \Rep{3} ^{\star}$ &
	 $\left( \begin{array}{c}
	\\
	\\
	\chi_{1}^{9} \, \chi_{3}^{2}
	\end{array} \right) \sim \Rep{3}$ &
	 $\left( \begin{array}{c}
	\chi_{1}^{8} \, \chi_{3}^{3}\\
	\\
	\\
	\end{array} \right) \sim \Rep{3}$\\
\hline
\end{tabular}
\begin{center}
\begin{minipage}[t]{22cm}
\caption[]{Higher-dimensional operators of $T_7$. We assume that the VEVs of the fields $\chi_i$
are of the form $\langle \chi_1 \rangle = \eta\, \Lambda$, 
$\braket{\chi_{2}}/\braket{\chi_{1}} \approx \epsilon ^4$, $\braket{\chi_{3}}/
\braket{\chi_{1}} \approx \epsilon ^2$ with $\eta \sim \mathcal{O}(1)$ and $\epsilon \approx 0.05$.
Note that in the cases in which the one-dimensional representation $\MoreRep{1}{1}$ is given
as a polynomial, {\it i.e.} for $n>3$, also similar polynomials forming the representations 
$\MoreRep{1}{2,3}$ exist leading to contributions of the same order in $\epsilon$ as the one transforming as 
$\MoreRep{1}{1}$. Note further the periodicity of the covariants in $n$, {\it e.g.} for $n=4$ one finds that 
the combination $\chi_{1}^{2} \chi_{2} \chi_{3}$ is the first component of a triplet $\Rep{3}^{\star}$
and similarly for $n=4+7=11$ the monomial $\chi_{1}^{2+7} \chi_{2} \chi_{3}=\chi_{1}^{9} \chi_{2} \chi_{3}$ 
belongs to the first component of a $\Rep{3}^{\star}$. When using this table to compute the contributions 
to the fermion masses, one has to take into account a factor $\eta^{n}<1$ for each
covariant of order $n$.
\label{tab:highdimopT7}}
\end{minipage}
\end{center}
\normalsize
\end{table}
\end{landscape}

\newpage
\thispagestyle{empty}
\begin{table}
\footnotesize
\begin{center}
\begin{tabular}{|c|c|c|}
\hline
Order in $\epsilon$ & Operator Structure & Transformation Property\\
\hline
$\Order{1}$ & $\chi_{3} ^{m} \, (\chi_{3} ^{\star}) ^{n-m} \,\, (m=0,...,n)$
& $\MoreRep{1}{1,2,3}$ for $(2 \, m -n) \mod 3=0$
\\
& & $\rm 3^{rd}$ comp. of $\MoreRep{3}{2}$ for $(2 \, m -n) \mod 3=1$
\\
& & $\rm 3^{rd}$ comp. of $\MoreRep{3}{1}$ for $(2 \, m -n) \mod 3=2$
\\
\hline
$\Order{\epsilon^{4}}$ &
$\chi_{3} ^{m} \, (\chi_{3} ^{\star})^{n-1-m} \, \chi_{1}
\,\, (m=0,...,n-1)$
& $\rm 1^{st}$ comp. of $\MoreRep{3}{2}$ for $(2 \, m -n +1) \mod 3=0$
\\
& & $\rm 2^{nd}$ comp. of $\MoreRep{3}{3}$ for $(2 \, m -n +1) \mod 3=1$
\\
& & $\rm 3^{rd}$ comp. of $\MoreRep{3}{8}$ for $(2 \, m -n +1) \mod 3=2$
\\
\cline{2-3}
& $\chi_{3} ^{m} \, (\chi_{3} ^{\star})^{n-1-m} \, \chi_{1}^{\star}
\,\, (m=0,...,n-1)$
& $\rm 1^{st}$ comp. of $\MoreRep{3}{1}$ for $(2 \, m -n +1) \mod 3=0$
\\
& & $\rm 3^{rd}$ comp. of $\MoreRep{3}{7}$ for $(2 \, m -n +1) \mod 3=1$
\\
& & $\rm 2^{nd}$ comp. of $\MoreRep{3}{4}$ for $(2 \, m -n +1) \mod 3=2$
\\
\hline
$\Order{\epsilon^{2}}$ &
$\chi_{3} ^{m} \, (\chi_{3} ^{\star})^{n-1-m} \, \chi_{2}
\,\, (m=0,...,n-1)$
& $\rm 2^{nd}$ comp. of $\MoreRep{3}{2}$ for $(2 \, m -n +1) \mod 3=0$
\\
& & $\rm 1^{st}$ comp. of $\MoreRep{3}{3}$ for $(2 \, m -n +1) \mod 3=1$
\\
& & $\rm 1^{st}$ comp. of $\MoreRep{3}{7}$ for $(2 \, m -n +1) \mod 3=2$
\\
\cline{2-3}
  &
$\chi_{3} ^{m} \, (\chi_{3} ^{\star})^{n-1-m} \, \chi_{2}^{\star}
\,\, (m=0,...,n-1)$
& $\rm 2^{nd}$ comp. of $\MoreRep{3}{1}$ for $(2 \, m -n +1) \mod 3=0$
\\
& & $\rm 1^{st}$ comp. of $\MoreRep{3}{8}$ for $(2 \, m -n +1) \mod 3=1$
\\
& & $\rm 1^{st}$ comp. of $\MoreRep{3}{4}$ for $(2 \, m -n +1) \mod 3=2$
\\
\hline
$\Order{\epsilon^{6}}$ &
$\chi_{3} ^{m} \, (\chi_{3} ^{\star})^{n-2-m} \, \chi_{1} \, \chi_{2}
\,\, (m=0,...,n-2)$
& $\rm 3^{rd}$ comp. of $\MoreRep{3}{3}$ for $(2 \, m -n +2) \mod 3=0$
\\
& & $\MoreRep{1}{4,5,6}$ for $(2 \, m -n +2) \mod 3=1$
\\
& & $\rm 3^{rd}$ comp. of $\MoreRep{3}{6}$ for $(2 \, m -n +2) \mod 3=2$
\\
\cline{2-3}
  & $\chi_{3} ^{m} \, (\chi_{3} ^{\star})^{n-2-m} \, \chi_{1}^{\star} \, \chi_{2} \,\, (m=0,...,n-2)$
& $\rm 2^{nd}$ comp. of $\MoreRep{3}{8}$ for $(2 \, m -n +2) \mod 3=0$
\\
& & $\rm 1^{st}$ comp. of $\MoreRep{3}{6}$ for $(2 \, m -n +2) \mod 3=1$
\\
& & $\rm 2^{nd}$ comp. of $\MoreRep{3}{5}$ for $(2 \, m -n +2) \mod 3=2$
\\
\cline{2-3}
& $\chi_{3} ^{m} \, (\chi_{3} ^{\star})^{n-2-m} \, \chi_{1} \, \chi_{2} ^{\star} \,\, (m=0,...,n-2)$
& $\rm 2^{nd}$ comp. of $\MoreRep{3}{7}$ for $(2 \, m -n +2) \mod 3 =0$
\\
& & $\rm 2^{nd}$ comp. of $\MoreRep{3}{6}$ for $(2 \, m -n +2) \mod 3 =1$
\\
& & $\rm 1^{st}$ comp. of $\MoreRep{3}{5}$ for $(2 \, m -n +2) \mod 3 =2$
\\
\cline{2-3}
& $\chi_{3} ^{m} \, (\chi_{3} ^{\star})^{n-2-m} \, \chi_{1}^{\star} \,
\chi_{2}^{\star} \,\, (m=0,...,n-2)$
& $\rm 3^{rd}$ comp. of $\MoreRep{3}{4}$ for $(2 \, m -n +2) \mod 3=0$
\\
& & $\rm 3^{rd}$ comp. of $\MoreRep{3}{5}$ for $(2 \, m -n +2) \mod 3=1$
\\
& &  $\MoreRep{1}{7,8,9}$ for $(2 \, m -n +2) \mod 3=2$
\\
\hline
$\Order{\epsilon^{4}}$ &
$\chi_{3} ^{m} \, (\chi_{3}^{\star})^{n-2-m} \, \chi_{2} ^{2} \,\,
(m=0,...,n-2)$
& $\rm 2^{nd}$ comp. of $\MoreRep{3}{1}$ for $(2 \, m -n +2) \mod 3=0$
\\
& & $\rm 1^{st}$ comp. of $\MoreRep{3}{8}$ for $(2 \, m -n +2) \mod 3=1$
\\
& & $\rm 1^{st}$ comp. of $\MoreRep{3}{4}$ for $(2 \, m -n +2) \mod 3=2$
\\
\cline{2-3}
& $\chi_{3} ^{m} \, (\chi_{3}^{\star})^{n-2-m} \, \chi_{2} \, \chi_{2}^{\star}
\,\, (m=0,...,n-2)$
& $\MoreRep{1}{1,2,3}$ for $(2 \, m -n +2) \mod 3=0$
\\
& & $\rm 3^{rd}$ comp. of $\MoreRep{3}{2}$ for $(2 \, m -n +2) \mod 3=1$
\\
& & $\rm 3^{rd}$ comp. of $\MoreRep{3}{1}$ for $(2 \, m -n +2) \mod 3=2$
\\
\cline{2-3}
& $\chi_{3} ^{m} \, (\chi_{3}^{\star})^{n-2-m} \, (\chi_{2}^{\star}) ^{2} \,\,
(m=0,...,n-2)$
& $\rm 2^{nd}$ comp. of $\MoreRep{3}{2}$ for $(2 \, m -n +2) \mod 3=0$
\\
& & $\rm 1^{st}$ comp. of $\MoreRep{3}{3}$ for $(2 \, m -n +2) \mod 3=1$
\\
& & $\rm 1^{st}$ comp. of $\MoreRep{3}{7}$ for $(2 \, m -n +2) \mod 3=2$
\\
\hline
$\Order{\epsilon^{6}}$ &
$\chi_{3} ^{m} \, (\chi_{3} ^{\star})^{n-3-m} \, \chi_{2} ^{3} \,\,
(m=0,...,n-3)$
& $\MoreRep{1}{1,2,3}$ for $(2 \, m -n) \mod 3=0$
\\
& & $\rm 3^{rd}$ comp. of $\MoreRep{3}{2}$ for $(2 \, m -n) \mod 3=1$
\\
& & $\rm 3^{rd}$ comp. of $\MoreRep{3}{1}$ for $(2 \, m -n) \mod 3=2$
\\
\cline{2-3}
& $\chi_{3} ^{m} \, (\chi_{3} ^{\star})^{n-3-m} \, \chi_{2} ^{2}
\, \chi_{2} ^{\star} \,\,
(m=0,...,n-3)$
& $\rm 2^{nd}$ comp. of $\MoreRep{3}{2}$ for $(2 \, m -n) \mod 3 =0$
\\
& & $\rm 1^{st}$ comp. of $\MoreRep{3}{3}$ for $(2 \, m -n) \mod 3 =1$
\\
& & $\rm 1^{st}$ comp. of $\MoreRep{3}{7}$ for $(2 \, m -n) \mod 3 =2$
\\
\cline{2-3}
& $\chi_{3} ^{m} \, (\chi_{3} ^{\star})^{n-3-m} \, \chi_{2} \,
(\chi_{2} ^{\star})^{2} \,\,
(m=0,...,n-3)$
& $\rm 2^{nd}$ comp. of $\MoreRep{3}{1}$ for $(2 \, m -n) \mod 3 =0$
\\
& & $\rm 1^{st}$ comp. of $\MoreRep{3}{8}$ for $(2 \, m -n) \mod 3 =1$
\\
& & $\rm 1^{st}$ comp. of $\MoreRep{3}{4}$ for $(2 \, m -n) \mod 3 =2$
\\
\cline{2-3}
& $\chi_{3} ^{m} \, (\chi_{3} ^{\star})^{n-3-m} \, (\chi_{2} ^{\star})^{3} \,\,
(m=0,...,n-3)$
& $\MoreRep{1}{1,2,3}$ for $(2 \, m -n) \mod 3=0$
\\
& & $\rm 3^{rd}$ comp. of $\MoreRep{3}{2}$ for $(2 \, m -n) \mod 3=1$
\\
& & $\rm 3^{rd}$ comp. of $\MoreRep{3}{1}$ for $(2 \, m -n) \mod 3=2$
\\
\hline
\end{tabular}
\end{center}
\begin{center}
\begin{minipage}[t]{17cm}
\caption[]{Higher-dimensional operators of $\Sigma(81)$.
We assume the vacuum structure $\braket{\chi_{1}}/\braket{\chi_{3}} \approx \epsilon ^4$,
$\braket{\chi_{2}}/\braket{\chi_{3}} \approx \epsilon ^2$, $\braket{\chi_3} = \eta \, \Lambda$
with $\eta \sim  \mathcal{O}(1)$ and $\epsilon \approx 0.05$.
Analogous to $T_7$, we can uniquely identify as which component of a three-dimensional representation
a certain monomial in the fields $\chi_i$ and $\chi_j^{\star}$ transforms by using the three elements
$\rm S_1=C^{2}$, $\rm S_2=A^{2} \, C^{2} \, A$ and $\rm S_3=A \, B^{2} \, C \, A^{2}$ of the group 
which are products of the generators $\rm A$, $\rm B$ and $\rm C$, 
see \Tabref{tab:genSigma81}.  
The resulting transformation properties are shown in the third column.
The number of operators with a certain transformation property is approximately $[\frac{n}{3}]$ for
larger values of $n$. 
\label{tab:highdimopSigma81}}
\normalsize
\end{minipage}
\end{center}
\end{table}

\end{document}